\documentclass[preprint,12pt]{aastex}

\usepackage{graphicx}
\usepackage{lscape}

\newcommand{\noprint}[1]{}
\newcommand{\figsetstart}{{\bf Fig. Set} }
\newcommand{\figsetend}{}
\newcommand{\figsetgrpstart}{}
\newcommand{\figsetgrpend}{}
\newcommand{\figsetnum}[1]{{\bf #1.}}
\newcommand{\figsettitle}[1]{ {\bf #1} }
\newcommand{\figsetgrpnum}[1]{\noprint{#1}}
\newcommand{\figsetgrptitle}[1]{\noprint{#1}}
\newcommand{\figsetplot}[1]{\noprint{#1}}
\newcommand{\figsetgrpnote}[1]{\noprint{#1}}

\begin{document}

\title{Census of blue stars in SDSS DR8}

\author{
Samantha Scibelli\altaffilmark{\ref{BHBL},\ref{RPI}},
Heidi Jo Newberg\altaffilmark{\ref{RPI}},
Jeffrey L. Carlin\altaffilmark{\ref{RPI}}, \&
Brian Yanny\altaffilmark{\ref{FNAL}}
}

\altaffiltext{1}{Burnt Hills-Ballston Lake High School\label{BHBL}}
\altaffiltext{2}{Dept. of Physics, Applied Physics and Astronomy, Rensselaer
Polytechnic Institute, 110 8th Street, Troy, NY 12180, heidi@rpi.edu\label{RPI}}
\altaffiltext{3}{Experimental Astrophysics Group, Fermi National Accelerator Laboratory,
PO Box 500, Batavia, IL 60510\label{FNAL}}

\shortauthors{Scibelli, Newberg, Carlin \& Yanny}

\begin{abstract}

We present a census of the 12,060 spectra of blue objects 
($(g-r)_0<-0.25$) in the Sloan Digital
Sky Survey (SDSS) Data Release 8 (DR8).    
As part of the data release, all of the spectra were 
cross-correlated with 48 template spectra of stars, galaxies and QSOs to determine the 
best match.  We compared the blue spectra by eye to the templates assigned in SDSS DR8.
10,856 of the objects matched their assigned template, 170 could not be classified due to low signal-to-noise (S/N), and 
1034 were given new classifications.  
We identify 7458 DA white dwarfs, 1145 DB white dwarfs, 273 rarer white dwarfs (including carbon, DZ, DQ, and magnetic), 294 subdwarf O stars, 648 subdwarf B stars, 679 blue horizontal branch stars, 1026 blue stragglers, 13 cataclysmic variables, 129 white dwarf - M dwarf binaries, 36 objects with spectra similar to DO white dwarfs, 179 QSOs, and 10 galaxies.  We provide two tables of these objects, sample spectra that match the templates, figures showing all of the spectra that were grouped by eye, and diagnostic plots that show the positions, colors, apparent magnitudes, proper motions, etc. for each classification.  Future surveys will be able to use templates similar to stars in each of the classes we identify to classify blue stars, including rare types, automatically.  

\end{abstract}


\section{Introduction}

In introductory astronomy classes, we teach the MK system \citep{1973ARA&A..11...29M} for classifying stars.
This system is fundamentally described by a set of prototype spectra on photographic 
plates, which can be compared with the spectrum of an unclassified star, imaged on a 
similar photographic plate \citep{2007AJ....134.1072C}.  In contrast,
modern stellar spectra are obtained with CCD cameras, and visualized as ``traces," 
which are plots of flux per unit wavelength as a function of wavelength.  In general, 
these traces are classified by comparing with a library of stellar spectra to 
determine which is the closest match \citep{1996A&AS..118..595V}.  The libraries should contain stellar spectra with 
a range of temperatures, surface gravities and metallicities that are either generated
from simulations and massaged to match the resolution and sensitivity of the observed 
spectrum, or contain spectra of stars previously obtained on the same spectrograph that have 
measured values for temperature, surface gravity, and metallicity.  Instead of 
classifying on the MK system, stars are grouped by derived properties.  Nonetheless, stars are still often referred to using the name of a star on the MK system that has a similar 
temperature and surface gravity.  

In this paper we discuss spectra extracted from Data Release Eight of the Sloan Digital Sky Survey (SDSS DR8; Aihara et al. 2011, York et al. 2000).
The SDSS provides uniform photometry in five magnitude passbands: $u, g, r, i,$ and $z$ (Krisciunas et al. 1998) over a quarter of the entire sky, primarily at high ($|b|>30^\circ$) latitudes.  Spectra of stars were obtained for purposes of calibration and to fill available fibers in the SDSS spectroscopic survey of galaxies and the SDSS II Legacy survey, which completed the initially proposed SDSS survey.  In addition, spectra of stars were obtained in the Sloan Extension for Galactic Understanding and Exploration (SEGUE; Yanny et al. 2009) survey: SEGUE-1 in SDSS II, and SEGUE-2 in SDSS III.  The SEGUE plates did not include objects selected as extragalactic objects; stars were the primary science targets.  The calibration stars and stars that filled available fibers were observed throughout the entire SDSS footprint, whereas the SEGUE spectra include deeper and more complete samples of stars over small areas, each covered by a bright and a faint survey `plate'.
Although the SDSS was primarily an extragalactic survey, it produced 
important results in many other areas of astronomy, including the identification of 
new types of stars \citep{1999ApJ...522L..61S}, discovery of Galactic halo substructure 
\citep{nyetal02}, and even the colors and statistics of asteroid families \citep{ivezicetal02}.

In the early days of the SDSS survey, the SDSS spectroscopic reduction team selected a set of sample star spectra from its first years of operation to serve as spectroscopic templates.  These were correlated  with all spectra of stars, for the purpose of radial velocity 
determination.   For determining radial velocities, it is better 
to have a template which closely matches the target spectrum, and thus template stars 
from a variety of spectral types were chosen.  This set of spectral templates 
consisted of 48 SDSS spectra with types ranging from O through L and T, including 
white dwarf stars.  However, all spectral types were not sampled, and the classification 
of the templates themselves is not rigorous. That is, these spectra are themselves
SDSS spectra, with $g > 15$, that are not directly classified using the Morgan Keenan 
system of MK standards.  When one shifts and correlates each template against the 
target spectrum, the best fit template (in a minimal chi-sq sense), also becomes 
the SDSS's reference classification for that particular target star, and 
is marked as such in the database, in the SUBCLASS field of 
the STAR table.

SDSS stars are generally classified neither from the templates nor 
using the MK system of OBAFGKM types.  The SDSS template spectra that 
are used in this paper are not in fact well known outside of the SDSS 
analysis team.  Instead stars are classified numerically by their 
effective temperature, metallicity, and surface gravity, as determined 
by comparison with stellar atmosphere models.  In SDSS DR8, these 
measurements are made with the Sloan Stellar Parameters Pipeline 
\citep{SSPP}.  Note however that stellar atmosphere models were not 
available for the hottest stars $T_{\rm eff} > 8500$ K (though see 
Eisenstein et al. 2006 for WDs), and thus the SDSS tabulated effective 
temperatures may not be reliable at the hottest stellar temperatures, 
including the stars investigated in this paper.  

Throughout this paper we will use colors that have been dereddened with 
the full line-of-sight reddening, as determined from \cite{sfd} maps, and indicated by the 0 subscript.  For main sequence B 
and A stars which are present in this sample, this is appropriate; however, many 
white dwarfs in this sample are close enough that they sit in front 
of or within the dust layer of the Galaxy (scale height $< 125~ \rm pc$), and thus their 
photometric $(u-g)_0$ and $(g-r)_0$ colors may be `overcorrected' so they appear to be bluer
than they actually are.  To compare colors of white dwarfs with
other works, it is recommended that the catalog stars be re-selected  from the SDSS database with 
non-dereddened colors.  For purposes of this paper, the exact value
of the reddening is not important, as the photometry is only used in selection of the blue star sample and in the color-color and color-magnitude diagrams showing the properties of the classified spectra.

In this paper, we present a census of all of the SDSS spectra bluer than $(g-r)_0<-0.25$.  This range was selected because very few stars are expected in this color range at the high Galactic latitudes probed by SDSS, and particularly in the $15<g_0<20.5$ magnitude range observed.  The absolute magnitude of an A0 V star (on the red end of our color selection) is $M_g= -0.1$, so the SDSS spectra probe distances of 10 to 130 kpc from the Sun for these stars.  The lowest latitude with a significant number of spectra is $b=20^\circ$, which would put even the closest A0 V stars 3.5 kpc above the Galactic plane, where one expects nearly all stars to be from the halo, where star formation and young main sequence stars are not expected.  To detect disk stars in this apparent magnitude range, one needs a stellar tracer that is fainter than $M_g=3$; at colors this blue, the only stars fainter than that are white dwarfs.  Therefore, this color range yields a high fraction of interesting and rare objects, including white dwarf stars in the disk and horizontal branch stars ($M_g\sim-0.7$) in the halo.

We start by comparing the 12,060 blue stars in SDSS DR8 with the 
42 original templates.  Stars that do not match their assigned templates, typically because
they are unusual stars not represented in the template library, are labeled as 
`misclassified.'  This comparison was done visually by one author (Scibelli) by determining
whether a spectral template overplotted on top of the observed stellar spectrum looked
like a match.  
In one category (CV), many of the spectra were a good match to the template, but only because the SDSS template-matching routine allowed scaling by a negative multiplicative factor so that the cataclysimic variable emission lines became absorption lines.  We also considered these stars to be bad matches.

We find that 90\% of these blue spectra are well matched to nine of the stellar templates and five templates for QSOs and galaxies.  An additional 1.4\% have low enough signal-to-noise that classification is not possible.  The remaining 8.6\% consist of additional categories of hot stars.

We do not claim to assign proper MK spectral types to any of the stars; in fact many of these stars may not exist in the set of bright standards used by Morgan and Keenan.  Instead we supply a set of templates that describe 90\% of the stars, and the rest we group into sets of similar stars.  We then discuss the likely identifications of the star groups, in part based on the types that have been assigned to members of these groups by other authors.

\section{SDSS Target Selection of Very Blue stars}

The SDSS chose very blue stars for spectroscopic followup in several ways.  The primary category for selecting Very Blue stars is the  HOTSTD category of
the SDSS and SDSS II Legacy survey.  In these surveys, hot standards and brown dwarfs were considered rare and valuable, so they were given to the tiling algorithm with the highest priority.  After the tiling algorithm placed all possible galaxies and candidate QSOs within the survey criteria, left over fibers were assigned to various categories of stars, serendipity objects, ROSAT sources, and other standards.  Because of the very high priority, a spectrum was taken of essentially all stellar objects with
$g_0 < 19, (u-g)_0 < 0$ and $(g-r)_0 < 0$ within the survey footprint.   
 Since this selection category uses dereddened magnitudes and colors, it will pick up 
some objects that are not intrinsically very blue, but that lie within or in front of the dust column and have thus been overcorrected by the application of the \citet{sfd} extinction value.
It should, on the other hand, contain spectra of all very 
blue objects so long as they are not too faint, and in this sense be 
an essentially complete sample of such objects.  

Note, however, that there are also many objects with $(g-r)_0 < -0.25$, but with $(u-g)_0 > 0$
which are not selected within the HOTSTD category.  Many of the stars with
these photometric colors were targeted by the SDSS and SDSS II Legacy category aimed at selecting blue horizontal branch stars (BHBs; STARBHB),
and the SEGUE category BHB/BS/A (SEGUE-2 has a similar BHB category; see Yanny et al. 2009, Tables 7 and 5).  This category is well populated in the SDSS Legacy survey of $>8000~\rm deg^2$ of sky, but not 100\% spectroscopically complete (that is not every star with these colors was targeted for a spectrum by SDSS Legacy and certainly not by SEGUE, since SEGUE covered a much smaller area).
Also, the legacy BHBs were required to have $(u-g)_0 > 0.8$.  Thus, there is a gap of very blue stars with $ 0 < (u-g)_0 < 0.8, (g-r)_0 < -0.25$ where the HOTSTD and the Legacy BHB categories do not cover.   The STARWHITEDWARF, SERENDIPBLUE $(u-g)_0 < 0.8, (g-r)_0 < -0.05$ fills this gap, though, again not with
100\% completeness, since not every SERENDIPBLUE candidate with the proper
photometric colors could be assigned a spectroscopic fiber.  
There are also a number of very blue stars which were spectroscopically targeted
by SDSS as QSOs, but which in fact turned out to be very blue stars.
The color cut for quasars requires $(u-g)_0 < 0.6$, which also helps fill
the gap redward of the HOTSTD $(u-g)_0 > 0$ cut and the BHB Legacy cut. 

We roughly estimate that for stellar (point) sources with $(g-r)_0 < -0.25, 14.5 < g_0 < 19$,
SDSS DR8 is 90\% spectroscopically complete for objects additionally with $(u-g)_0 < 0$ (HOTSTD), 85\% complete for $0 < (u-g)_0  < 0.6$ (failed QSOs), 90\% complete for $ 0.6 < (u-g)_0 < 0.8$ (SERENDIPBLUE), and 95\% complete for $0.8 < (u-g)_0 < 1.8$ (BHBs).  These numbers were derived by querying SDSS DR8 for stellar objects with photometry in these color ranges, and then determining for what fraction of these objects spectra had been obtained.

This high completeness suggests that our classification of very blue objects from SDSS DR8 is very representative of all such objects in the Galaxy at
high Galactic latitude $|b| > 30^\circ$, at these magnitude limits.  
Note that because of the Galactic latitude limit and the saturation limit 
of SDSS $g\sim 14.5$, we will not be sampling completely O and B main sequence stars and may be completely missing whole categories of blue stars such as Wolf-Rayet stars.

\section{Data Selection and Template Matching}

We selected all stars with spectra from SDSS DR8 with an extinction-corrected color cut 
of $(g-r)_0 < -0.25$.  These stars were selected using CasJobs, from the 
plateX table and star and specobj views available on the SDSS SkyServer Schema browser, 
using the query in Table 1.  
The query matches each object's ID number from the star view to its ID number in the 
specobj view as well as each object's plate number in specobj and plateX. This verifies 
that each object is unique in the dataset so there are no repeats. 
Note that the extinction-corrected magnitude for $u, g$ and $r$ is cut at 22; spectra of objects
fainter than 22 are sky fibers. 
Also the selections $g > -9999$ and $g_{\rm error} > -1000$ are used to remove objects with 
no measured $g$ magnitude. The $ugr$-band extinction cut ($A_u, A_g$, and $A_r < 2$) is used to remove objects in heavily reddened areas.  Note that this query uses the `star' table from SDSS, which includes only point-like sources from the photometric database. However, no restrictions were placed on the spectroscopic class, so that QSOs and compact galaxies are present in the sample.
We included objects classified as extragalactic sources because there were
very few QSOs and galaxies with colors this blue, and a significant fraction of those were
misclassified.  A parameter with a ``clean'' flag equal to one ensures that the objects in the data do not
have `saturated' or `edge' or `interpolated' flags that often lead to bad photometric
measurements.

In total there were 12,113 objects selected from the query.  Of those objects, 39 did not have spectra in DR8, meaning that for some reason the data from the fiber was not processed through the pipeline.  Possibly, there was not useful data in that fiber.  The remaining 12,074 spectra were examined by eye. 

\begin{figure}
\includegraphics[width=5.0in]{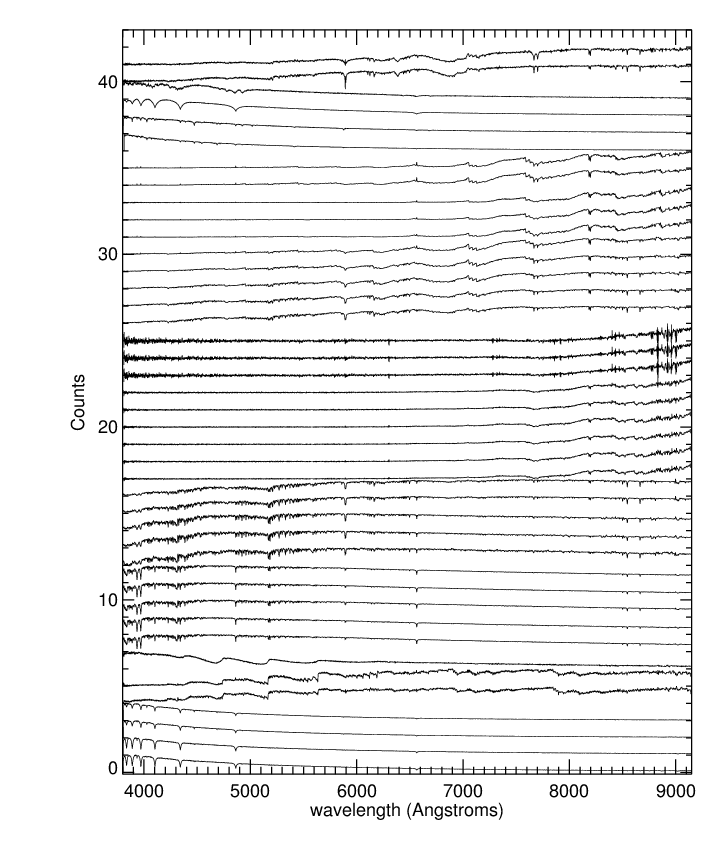}
\caption[stellartemplates] {
\footnotesize
The 42 stellar templates to which all observed SDSS DR8 spectra were matched.  They 
are scaled so the maximum flux is one, and shifted by the template number.  Templates 1-4 are
hot stars (A0, A0p, B6 and B9).  Templates 5-7 are carbon stars (carbon, carbon with lines, 
and carbon white dwarf).  Templates 8-17 are cool stars (F2, F5, F9, G0, G2, G5, K1, K3, K5,
K7).  Templates 18-26 are brown dwarfs (L0, L1, L2, L3, L4, L5, L5.5, L9, T2).  Templates
27-36 are M stars (M0, M1, M2, M3, M4, M7, M8, M9, M5, M6; note that the last two are out of
temperature sequence).  Templates 37-38 are very hot
stars (O, OB).  Templates 39-40 are white dwarfs (DA WD, magnetic WD).  Templates 41-42
are M main sequence stars (M0V, M2V).
}\label{stellartemplates}
\end{figure}

\begin{figure}
\includegraphics[width=5.0in]{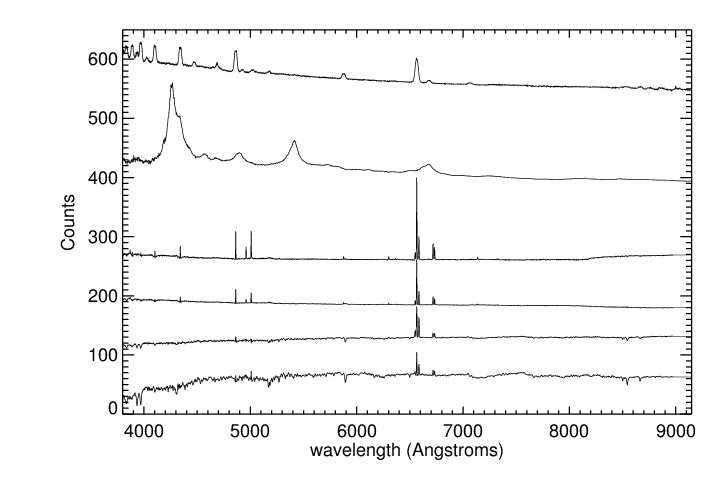}
\caption[galtemplates] {
\footnotesize
The CV, QSO and galaxy templates used by the SDSS pipeline. The bottom four spectra are galaxies, the fifth from the bottom is a QSO spectrum, and the upper trace is a CV.
}\label{galtemplates}
\end{figure}

SDSS DR8 includes a cross-correlation with 48 templates to estimate the type of object represented by each spectrum.
The template matches are used to identify stars, QSOs, galaxies, and
cataclysmic variables (CVs) in the database.  The templates shown in Figure~\ref{stellartemplates} include 42 sub-classes of stars.  They also include four galaxy templates, one 
QSO, and one cataclysmic variable template, as shown
in Figure~\ref{galtemplates}.  The four galaxy templates span a range of emission line
strength, compared to calcium absorption.  

The template names given here, in the figures, and in the table are the labels in the SDSS databases, and in some cases are not the best description of the spectral type.  These spectra are all from high Galactic latitude and fainter than $g=15$, so one would not expect to find {\it any} main sequence O stars in the dataset.  By comparison to classification spectra in Gray and Corbally (2009), we surmise that the O template is a late subdwarf O star that is both helium-rich and carbon-rich.  The stars that have temperatures of O and B stars should be primarily (or all) white dwarf or subdwarf stars.

The templates in general are an odd set; there are a large number of templates for M stars and brown 
dwarfs, and very few hot stars.  They are stars selected from SDSS spectra to span a range of temperature.  They therefore represent intrinsically bright objects that are common in the Milky Way halo and intrinsically very faint stars in the local solar neighborhood.  In most cases no distinction is made between dwarf and giant stars; the classification is primarily in temperature.  In some cases the 
templates are poorly named even in temperature.  For example, the B templates are labeled B6 and B9, but the 
templates actually represent hotter B stars that are most likely subdwarfs.  There does not appear to 
be a template for cooler B stars in the set; 62 of the
stars that best matched the hot B templates did not match well because they were cooler B 
stars than the templates.  None of the other hot star templates seem to be rigorously checked
for accuracy of class (OB is anyway not a well defined type, and as we will see later selects white dwarf stars).  The L5.5, L9, and T2 templates are dominated by missubtracted night sky lines, since these stars have very little optical flux.  Therefore, many of the low signal-to-noise spectra, which are also often dominated by night sky lines, are matched to these templates.  In this paper, we 
accept each of the templates as representative of a class of stars that is represented in
the SDSS data, but the named type may not match the template label.

In order 
to look through the cataloged stars' spectra, the DR8 Bulk Search website was used (note that the DR8 website used for this paper is no longer available, but instead DR10 has been made available at http://dr10.sdss3.org/bulkSpectra/). Here, the spectra of the stars were displayed in 
black, with the best fit template plotted on top in red.  Any unusual-looking spectra that 
did not match the template assigned by the SDSS (or had extra features such as absorption 
lines that were not in the template spectrum) were identified and marked on the CSV file 
as well as manually on paper for later analysis. 

Our method of examining each spectrum by eye was time-consuming.  If an automated measure of the goodness of fit (for example $\chi^2$) were available, it might have helped to guide us to examine only the spectra which were numerically poor fits to the templates.  However, it is difficult to match the ability of a trained person.  Some spectra had poor sky subtraction or other data issues that could be ignored by a person but would not have been ignored by an automated algorithm.  Similarly, people are better at identifying clear departures from the template even if the spectrum is in general quite noisy.  The drawback of determining by eye whether two spectra match is that it is somewhat subjective and two people might make different decisions in some of the cases.  Besides Ms. Scibelli, other students and co-authors examined a subset of the spectra, but we decided that it was best to use the determinations of a single individual.

Some of the spectra with blue photometry turned out to have spectra indicating a redder star.
We discovered that these stars had flags, like ``DEBLENDED AT EDGE," which means that the
object was close enough to the edge of the frame that the deblending process is uncertain.
Table 2 gives the list of objects with bad photometry, that should not have been included in the
blue sample.

The ``Total" column in Table 3 gives a breakdown of the spectral types assigned to the remaining 12,060 stellar spectra, by template type.  In the Table 3, we only break down the spectral types by letter; subtypes such as K1, K3, K5, and K7 are combined.

Visual inspection of the actual spectra plotted over the assigned template indicated that 10,839 of the spectra match the templates well.  An additional 17 spectra differed from the assigned templates only due to data quality issues.  Some had bad sky subtraction, some had missing data in the spectrum (in some cases due to being observed by the bad fiber number 321).  Seven of the spectra appeared to have a flatfielding error near H$\delta$; these spectra were all observed with fiber number close to 81.  Since the parts of these spectra with good data were also good matches to the assigned templates, we included these as good matches, for a total of 10,856 good matches out of 12,060 spectra (90\%).

Of the remaining 1204 spectra (10\% of the original sample), 170 (1.4\% of the original sample) had spectra that were so noisy that it was not possible to assign a spectral type.  The templates that were assigned to these objects (QSO, Galaxy,
L, and T) are the templates that seemed to draw featureless and unusual spectra.  It is likely that these are also misclassified, but it is not possible to tell in which category they should be placed.  This group contains most of the spectra fainter than $g_0=20$, but also includes some objects as bright as $g_0=18.5$.  Many of these stars are classified as white dwarfs of various types in other catalogs, and we will show that this is consistent with the large proper motions for these stars.

The remaining 1034 stars were divided into ten groups of similar spectra.  We have named these groups to suggest our opinion about the type of object these spectra represent.

Table 3 shows the breakdown of how the 12,060 blue
spectra were classified by template matching in the SDSS. Because all of the stars in the sample are blue, as determined from $g-r$ color, there are no actual F, K, M, L, or T stars in the sample, even though some of the spectra were matched to these templates. 

Ninety percent (10,856 objects) of the blue sources
were well fit by the following templates: O (220), OB (848), B6 (209), B9 (365), A0 (1026), A0p (679), 
CV (7), carbon WD (9), DA WD (7246), magnetic WD (58), QSO (176), or galaxy (13).

Though the templates might not be well named, nonetheless 90\% of the SDSS spectra of blue objects are good matches to ten stellar spectra and three spectra of extragalactic objects.  In order to correctly identify the rarer blue stars, additional templates should be added to the classification set.  The next two sections describe the range of spectra that are, and are not, well matched to the SDSS templates.

\section{Stars that match the templates}

The 10,856 stars that match their templates are listed in Table 4. We provide the Equatorial position of each object; the SDSS Plate, modified Julian date (MJD), and Fiber; the SDSS class and template (subclass) that was matched to the spectrum; the signal-to-noise tabulated in the SDSS database; the SDSS $g_0$ magnitude and $(u-g)_0$ and $(g-r)_0$ colors; the SDSS proper motions in Galactic coordinates in units of mas/yr; and the classification of the same SDSS spectrum from three other catalogs, including white dwarfs \citep{2013yCat..22040005K}, white dwarfs and subdwarfs \citep{2006ApJS..167...40E}, and cataclysmic variables \citep{2011AJ....142..181S}.  The ten stellar templates that match blue stellar spectra in SDSS are named: O, OB, B6, B9, A0, A0p, CV, Carbon WD, WD, WD Magnetic.  Apparently, these spectra are representative of SDSS hot star spectra, since they span the range of properties for 90\% of the stars.  

In this section, we explore the nature of these representative stars, many of which have temperatures and absorption lines similar to stars with MK classifications of O, B, and A.  It is important to point out that these SDSS spectra are largely at high Galactic latitude and the stars are fainter than $g_0>14.5$.  Therefore we do not expect to find {\it any} young main sequence stars of type O, B, or A.  Main sequence disk stars of these temperatures are too close, and therefore have apparent magnitudes too bright for the SDSS survey.  We do not expect to find young stars in the halo population.  However, some of the A stars could be halo blue stragglers.  

We present representative spectra for the galaxies and QSOs in Figure 3, but do not discuss them in any further detail.  Figures 4, 5, 6, 7, 8, and 9 show the positions of the stars in Celestial coordinates, the $g_0$ vs $(g-r)_0$ H-R diagrams, $ugr$ color-color plots, the SDSS proper motions, the reduced proper motion diagrams ($H_g \equiv g_0 + 5 + 5 \log{\mu}$), and the distribution of signal-to-noise ratio.  We will use these figures to determine the nature of the stars associated with each spectral template.  Representative spectra for the objects that match the templates are given in Figure 3.

\subsection{Disk White Dwarfs: WD, WD Magnetic, Carbon WD, OB, and CV}

By far the largest fraction of blue stars are DA white dwarfs (7246 stars), that match the WD template.  These stars have very broad Balmer absorption lines.  

The WD Magnetic template is a DA WD star with multiple Balmer absorption lines because of Zeeman splitting of the energy levels of the electrons in the hydrogen atoms.  This template matches an additional 58 stars.  The Carbon WD template matches an additional nine stellar spectra.

The H-R diagram in Figure 5 and the $ugr$ color-color diagram in Figure 6 show that the $g_0$ magnitudes and $(g-r)_0$ colors of these three types of dwarfs are similar, but at the same $(g-r)_0$ color, magnetic white dwarfs are bluer than regular DA WDs in $(u-g)_0$, and Carbon WDs are bluer still in $(u-g)_0$.  Although these spectra were classified by absorption lines in the spectra, the identified types clearly separate in color.

The proper motion diagram in Figure 7 gives us confidence that these three white dwarf templates are in fact representative of white dwarf stars, since the panels for all three of these types show a group with large dispersion in the proper motions, as expected for nearby stars.  In addition, we see that both the ``OB" stars and the CV stars have similarly high proper motions.  

Since the apparent magnitude distribution of the OB stars and the CV stars is roughly similar to the WD stars, we know that the absolute magnitudes of OB, CV, and WD stars are similar.  Cataclysmic variable (CV) stars are mass transfer systems onto a white dwarf.  If the total intrisic brightness of the combined system is similar to a white dwarf, then the binary must be a star of similar apparent magnitude, which suggests that system is a white dwarf plus an M dwarf.  We might preferentially select systems where the companion is faint by selecting only blue stars.  All of these CV stars are identified in the \citet{2011AJ....142..181S} catalog of 285 CVs with spectra in SDSS.

The very high proper motions and faint apparent magnitudes of the stars that match the OB template tell us that these stars must also be white dwarfs.  The helium I lines indicate that these white dwarfs are of type DB.  We will see in the next section that many of the apparently brighter stars that appear to also be DB white dwarfs were marked as not matching the OB template; this will be discussed when we analyze those stars.

The DA white dwarfs and DB stars that match the OB template are observed in Figure 4 to cluster at low Galactic latitude, as one would expect for a disk population.  There are a much smaller number of CVs, but these are also clustered towards low Galactic latitude.

\subsection{Intrinsically Brighter Stars: O, B6, B9, A0, and A0p}

The proper motion diagrams in Figure 7 show that the stars that match the templates O, B6, B9, A0, and A0p have dispersions in their measured proper motions consistent with the measurement errors.  The proper motion distribution is similar to that of the QSOs, which are too distant to exhibit measurable tangential velocities.  The actual distribution of proper motions for these stars must have a sigma smaller than 4 mas/year.  If these are halo stars with a velocity distribution of 100 km/s, then the distances to these stars must be greater than 5 kpc (distance modulus greater than 13.6).  If they are disk stars, the velocity dispersion is smaller, resulting in a larger minimum distance, which is not consistent with their being disk stars.  We therefore conclude that these stellar populations must have an absolute magnitude brighter than turnoff stars, which have $M_{g_0}=4.2$.

The A0 stars are primarily at the very red end of our selection box, and at $(g-r)_0=-0.25$ they are very slightly redder in $(u-g)_0$ than the A0p stars.  The A0p stars also tend towards the red end of our selection, though not as strongly, and the reduced proper motion diagram suggests that the stars are intrinsically dimmer as the stars become bluer.  A comparison of the color-color diagram for separating blue straggler stars from lower surface gravity BHB stars (Yanny et al. 2000, Figure 10) suggests that the A0 stars could be the very blue tip of the blue straggler population and the A0p stars are on the very blue end of the ``horizontal branch."  The horizontal branch actually turns down blueward of $(g-r)_0<-0.3$ (the bluer stars are intrinsically fainter), which is consistent with our reduced proper motion results (e.g., see Smolinski et al. 2011, Figures 2 \& 3).  Note that these stars are on the extremely blue end of the horizontal branch.  In fact, the method of template convolution appears to be a great method for separating blue stars by surface gravity.

In the color-color diagram of Figure 6, we see that the O, B9, B6, and the bluer stars in the A0p selection form a sequence from bluer to redder in both $(u-g)_0$ and $(g-r)_0$.  This temperature sequence is consistent with the differences in the spectral absorption lines for these stars.

The only signficant halo population of very hot stars that are not white dwarfs are the subdwarfs (sdO and sdB), which are reviewed in \citet{2009ARA&A..47..211H}.  These stars are core helium-burning, or possibly even more evolved, stars at the blue end of the horizontal branch.  Subdwarf O stars are a more heterogeneous group, and include stars that have evolved past the red giant branch, past the horizontal branch, or even past the asymptotic giant branch.  Figure 1 of \citet{2009ARA&A..47..211H} suggests that sdO stars might be intrinsically brighter than sdB stars, which is consistent with our data; the apparent magnitudes of the stars of type ``O" cut off at about $g_0=19$, whereas the stars of type ``B9" are found all the way to the limit of the spectral survey, and are much more numerous at $g_0=19$, where the ``O" stars are no longer found.  This suggests that we are seeing the sdO star population to the limits of the spatial extent of the halo.  Note also that there are many more B stars with measurable proper motions even though there are almost twice as many A0 stars as B stars in the sample.

\section{Additional Categories of Blue Stars}

Our intent in classification was to group stars that look like each other together.  We identified ten separate groups of stars, a few of which have subclasses.  Because many of these star groups looked like stars in the published literature, we felt it was less confusing to name them by the type of star that they most likely represent.  However, in many cases the literature is unclear on classification or not enough information is available for these stars to make a fine judgement between different types of physical objects that have similar spectral features.  Although all of these objects should be investigated in greater detail to make a definitive classification, we have done our best to represent and count all of the different types of spectra of blue objects in the SDSS.

The sorting of the stars into categories was done by Newberg and Scibelli, by looking at the spectra by eye.  We were guided by sample spectra of blue objects, but in the end we separated the categories by similar spectral features alone, and named the categories by the most similar type from the literature.  We iterated many times through the categories to verify that the spectra were as uniform as possible.  The lines were identified from the NIST Atomic Spectra Database \citep{NIST} or from \citet{graycorbally}. Table 1 of \citet{1999ApJS..121....1M} gives a nice summary of the classification standards for white dwarfs.

Classifying spectra by eye is by its nature a subjective process.  Although the initial classifications were done by Scibelli, the final classification of the entire set of spectra was done by Newberg, so the classifications are consistent with the determinations of one person.  Some of the spectra, and in particular the fainter spectra with higher noise, had uncertain classification.  Although this would be true whether the classifications were automated or done by hand as these were, the classifications done by automated algorithms would be more repeatable.  In order to make the most consistent classifications possible, Newberg identified the essential and most prominent features of each spectral type that were then used to make the determination of type.  These prominent features are listed below in the subsections describing the individual types.

The ten categories of misclassified stars are given the suggestive titles; DA white dwarfs (DA), DB white dwarfs (DB), DO white dwarfs (DO), DZ white dwarfs (DZ), 
unusual white dwarf (UnuWD), `Binary Stars' (Bin), cataclysmic variable stars (CV), cool B stars (B), Hot stars (Hot), and stars with featureless blue spectra (Feat).  Many of the spectra fainter than $g_0>19.5$ could not be identified, and were therefore placed in the LowSN category.  These are preferentially redder objects, and typically have an S/N measurement of less than five.  
Table 3 shows the number of stars
in each category, and the templates that were originally (incorrectly) matched to the spectra.  

The DA white dwarfs are divided into three types, including white dwarfs with narrow lines (candidate low mass white dwarfs) and white dwarfs with broad Balmer absorption plus HeII lines.  The ``Hot" category is divided into three subtypes based on the types of absorption lines
identified.  We identify one subclass of ``UnuWD" stars (DQ) that was particulary uniform within this designation; there are many other stars that should be divided into subclasses in future studies. The B stars are also divided into two categories: B and ``cool B.'' 

The misclassified spectra are listed in Table 5, which includes information about each of the 1204 stars, including the stars with low signal-to-noise, that did not match one of the thirteen blue SDSS templates.  We provide the Equatorial position of each object; the SDSS Plate, modified Julian date (MJD), and Fiber; the SDSS template that was matched to the spectrum; the signal-to-noise tabulated in the SDSS database; the SDSS $g_0$ magnitude and $(u-g)_0$ and $(g-r)_0$ colors; the proper motions in Galactic coordinates in units of mas/yr; a measure of the $r_0$ variability
(dr); the classification assigned in this paper (NewCL); and the classifications of the same SDSS spectra from three other catalogs: white dwarfs \citep{2013yCat..22040005K}, white dwarfs and subdwarfs \citep{2006ApJS..167...40E}, and cataclysmic variables \citep{2011AJ....142..181S}.  The variability ($\Delta r_{\rm max}$) is the difference between the faintest and brightest $r_0$  magnitude of all of the photometric measurements in SDSS DR8. The number in parentheses is the number of observations in SDSS DR8.

In addition to the table, we include traces of each of the 1204 spectra, grouped by classification, in Figures 18-28.  These figures can be used to evaluate the variation within a classification, and are useful for identifying extremely unusual objects.  In order to interpret our classifications, we have made diagnostic plots (Figures 10-15) for each of the ten types, the low S/N spectra, and all spectra together, that are similar to the diagnostic plots for the stars that {\it did} match their templates.  Figures 16 and 17 show a measure of the photometric variability of the stars of each identified spectral type.

Figure 10 shows the distribution on the sky for each type of star.  The DA, DB, and unusual white dwarf stars are clearly clustered towards the Galactic plane.  The DA white dwarfs are not only clustered at low Galactic latitudes towards the anticenter, but also at slightly higher Galactic latitudes towards the Galactic center.  This is probably because they are typically more distant in our sample, and therefore we see the larger number of stars towards the center of the Milky Way.

Figure 11 shows an H-R diagram for each type.  The H-R diagram for ALL spectra shows the systematics of the target selection process.  For example few stars around $g_0=19$ were observed; the Legacy plates and bright SEGUE plates had a limiting magnitude of about 18.5, and fainter SEGUE plates had a limiting magnitude of about 20.5 (though the limits and filters used vary by target selection category).  Since fainter stars typically dominate the sample in any magnitude range selected, the magnitudes of the stars observed favored the plate limits - particularly for the fainter plates.

Figure 12 shows $ugr$ color-color diagrams by classification type.  It is encouraging that the different types, classified entirely by inspection of the spectra and not by their photometry, occupy different areas of the color-color diagram.  The possible exception is that LowSN spectra have very similar colors to the featureless blue spectra (which naturally will be very difficult to identify), but have fainter apparent magnitudes, as shown in Figure 11.

The proper motions of each category are shown in Figure 13, and the reduced proper motion diagrams are shown in Figure 14.  These show that the hotter stars (DO, B, and Hot) are typically more distant, as most of the proper motions are within errors.  However, a fraction of the stars in these categories do appear to have significant proper motions.  This will be discussed further in the respective sections.  The intrinsically dimmest objects are the unusual white dwarfs, the featureless blue spectra, and the low S/N objects.  By comparing the scatter in the proper motion and H-R diagrams, we see that the UnuWDs are closer and have brighter apparent magnitudes than those of the featureless blue spectra, which are in turn closer and have brighter apparent magnitudes than the low S/N spectra.  So there is a possibility that the objects in these categories are intrinsically similar, but at different distances.

The characteristics of the ten subsets of misclassified stars except lowSN spectra (for which we have little additional information) are described below.

\subsection{DA White Dwarf Stars}

There were only three SDSS template spectra for white dwarfs: a
DA white dwarf (though not a particularly hot DA white dwarf), a magnetic white dwarf, and a carbon white dwarf.
In cases where the DA white dwarf was hotter, and therefore the absorption lines were not
as deep, the template matching instead selected a B or O star, or selected a QSO spectrum
with a negative flux (a negative scaling factor seems to have been allowed in the SDSS template matching routine) so that the QSO 
emission lines became the white dwarf absorption lines.  The spectral templates for L5.5, L9, and T2 were dominated by bad night sky line subtraction, so these were often chosen to match low signal-to-noise white dwarf spectra.

The 212 stars that did not match the DA white dwarf template, the A0 template, or the A0p template, and whose primary identifying features were wide Balmer absorption lines, were put in this DA white dwarf category.  Most of the ``misclassified"
objects in this category were of the hot DA white dwarf variety, with much shallower absorption lines than the DA white dwarf template in SDSS.  Some of the stars, in addition to apparently broad Balmer absorption lines, also showed evidence of He II absorption, so they were classified as ``DA + HeII."  Two of the stars had deep Balmer absorption lines, but the lines were narrower than we expect for their color.  These were subclassified as ``DA Low Mass."  

The vast majority of the stars we classified as DA WDs were also classified by \citet{2013yCat..22040005K} as DA white dwarfs, though they broke the types down into a larger number of categories.  Many of the previously identified stars in our ``DA+HeII" category were classified by \citet{2006ApJS..167...40E} as DAO.  All we can say is that to our eyes, all of the stars in this category appeared to have broad Balmer absorption, the Low Mass subcategory had narrower lines for the depth, but not as narrow as the A0 star templates, and the DA+HeII stars had He II absorption lines.  We also note that the DA+HeII stars all have signal-to-noise greater than ten; many of the stars listed as ``DA" might have been classified as DA+HeII in a higher signal-to-noise spectrum.  We note that neither of the two candidate low mass DA WDs was previously classified as a white dwarf.  However, the proper motions of these stars suggest our classification is correct; proper motions were only examined after the identifications were made.

The color-magnitude diagram for DA white dwarfs shows that most DA white dwarfs have $-0.6<(g-r)_0<-0.45$.  However, particularly at the faint end there are many apparently cooler objects classified as white dwarf stars.  There is apparently a very uniform group of hot objects that define the DA white dwarfs (or at least the DA white dwarfs that do not match the SDSS template spectrum).  A small fraction of the stars show variability, which is likely the result of an undetected companion.

\subsection{DB White Dwarf Stars}

The DB WD spectra show broad absorption lines of neutral helium (He I).  The strongest lines are at
3889 \AA, 4026 \AA, 4471 \AA, 4921 \AA, and 5876 \AA.  The lines at 4026 \AA~and 4471 \AA~are particularly strong. In addition, the spectra also show absorption at 3964 \AA, 4120 \AA, 4388 \AA, 4713 \AA, 5016 \AA, and 6678 \AA.  As a general rule, the spectra in this group have broader absorption lines than the spectra that matched the ``OB" template, though there are a variety of ways the spectrum could have been different from the template.

The DB WDs occupy a narrow range of $ugr$ colors, which may be truncated a bit on the red end due to our selection of only stars bluer than $(g-r)_0<-0.25$.  They can apparently be identified at the faintest magnitudes in the survey, to which their population extends.  Like DA WDs, they are found over all surveyed Galactic latitudes, though they are found preferentially at low Galactic latitude.  The reduced proper motion diagrams show these stars as redder and fainter than DA WDs.  As a group, they show no special propensity for variability, and in this regard seem similar to the DA WDs. 

As was true for the DA WDs, the vast majority of the stars we classified as DA WDs were also classified by \citet{2013yCat..22040005K} as DB white dwarfs of some type.

\subsection{DO White Dwarf Stars}

Any star whose strongest absorption line is He II at 4686 \AA \citep{2004A&A...417.1093K} was classified as a DO WD.  Some of the spectra also have He II lines at 4861 \AA, 5412 \AA, and 6560 \AA, and/or He I at 4471 \AA.  Since there was no DO white dwarf in the SDSS template set, these stars usually matched the O star template.  Only about a third of these stars have detected proper motions, which leads us to believe that a large fraction of the stars in this category are not actually white dwarfs.

The distribution of these in the sky is also unusual.  Very few are found at high Galactic latitudes, and a disproportionate number are located in the south Galactic cap.  The sky distribution appears far clumpier than we would expect from a random distribution of stars.  

To show this, we first compare the distribution in Galactic latitude of the OB stars that match their templates with the distribution in Galactic latitude of the DB stars that did not match their templates.  At latitudes of $b<0^\circ, 0^\circ<b<30^\circ, 30^\circ<b<60^\circ,60^\circ<b$, we see 153, 96, 422, 177 OB stars, respectively.  At the same set of Galactic latitudes we see 44, 47, 147, 59 DB stars, respectively. The percentages of each type of star in each range is similar: $18.0 \pm 1.3\%, 11.3 \pm 1.1\%, 49.8 \pm 1.7\%$ and $20.9 \pm 1.4\%$ for OB stars versus $14.8 \pm 2.1\%, 15.8 \pm 2.1\%, 49.5 \pm 2.9\%,$ and $19.9 \pm 2.3\%$ for DB stars.  The $\chi^2$ test results in a $\chi^2$ value of 5.02, with three degrees of freedom.  This is equivalent to a p-value of 0.17.  This is not sufficient to reject the null hypothesis that these two sets of stars are distributed similarly in Galactic latitude.  Since we expect that both sets of stars represent DB white dwarfs, we take this as evidence that these two distributions are in fact the same.

We therefore compare the distribution of candidate DO white dwarf stars to the sum of the counts of DB and DO stars in each Galactic latitude range.  We see 11, 7, 5, and 3 DO stars at Galactic latitudes of $b<0^\circ, 0^\circ<b<30^\circ, 30^\circ<b<60^\circ,$ and $60^\circ<b$, respectively.  The percentages in each latitude bin are: $42.3 \pm 9.7$ \%, $26.9 \pm 8.7$ \%, $19.2 \pm 7.7$\%, and $11.5 \pm 6.3$\% .  The $\chi^2$ test results in a $\chi^2$ value of 19.0, with three degrees of freedom.  This is equivalent to a p-value of 0.0003, which is sufficient to rule out the hypothesis that the DO stars are distributed the same way the DB stars are distributed.  This is somewhat surprising because one would assume that DO WDs are of similar absolute magnitude to DB white dwarfs.  The finding that $42 \pm 10$\% of the DO stars are in the southern hemisphere, when one might more reasonably expect only $17 \pm 1$\%, is unusual.

The $(g-r)_0$ colors of DO WD stars are similar to DA WD stars, but the $(u-g)_0$ color is significantly bluer.  Our sample includes few DO WD objects fainter than 19$^{\rm th}$ magnitude, indicating either that we have hit the distant limit for this population, or that we are unable to detect them at low S/N.  It is possible that we do not identify fainter DO WD because the absorption lines are weak; there are a few stars in the ``Featureless blue objects" category that have the same $ugr$ colors, and it is possible that there are two populations of featureless objects apparent in the proper motion diagrams: one with low proper motions like the DO stars and one with much higher proper motions expected from a nearer stellar population.  Three of the brightest stars in this category show possible variability, but the remainder of the population shows no significant variability.

Most of the stars we classify as DO are also classified by \citet{2013yCat..22040005K} as DO or PG1159 types.  PG1159 stars are the pre-degenerate hot central stars of planetary nebulae, which will eventually lose mass, cool and become DO WD stars.  Since some could be fusing helium, they can be substantially brighter than DO white dwarfs.  Upon re-examination of our DO WD spectra, the eight PG1159 stars are immediately identifiable from their double absorption lines of He II plus C IV in the 4650-90 \AA~region.  The inclusion of these stars in our DO classification partially explains the small proper motion sample, but does not explain all of them.  We leave our classifications as they are to preserve the independence of our results.

\subsection{DZ White Dwarf Stars}

We have identified seven objects with very blue spectra, but very broad and deep Ca H\&K absorption lines at 3968 \AA~and 3933 \AA.  These match the DZ white dwarf templates in \citet{1990ApJS...72..707S}, \citet{2006ApJS..167...40E}, and \citet{2007ApJ...663.1291D}.  Four of these stars also have broad He I lines like DB WDs, and two have evidence of Balmer lines.  The third has extremely deep Ca H\&K, and does not show any hydrogen or helium absorption.  Six of these stars have some sort of DZ (including DB\_DBZ) designation in \citet{2013yCat..22040005K}, and one is classified as DB.  The DZ WDs have quite high proper motions, and are likely some of the fainter white dwarfs in the blue survey. 

Note that all seven of the stars in our DZ category have a $(g-r)_0$ color that is bluer than any of the DZ stars in \citet{2007ApJ...663.1291D}, so these are an unusual set of DZ stars.  None of the DZ stars show any evidence of variability.

\subsection{Unusual White Dwarf Stars}

Any star that had broad absorption lines and wiggles in the spectrum that are associated with a white dwarf, but the absorption lines are not at the locations expected for DA, DB, or DO WD stars were put into the unusual white dwarf category.  In addition, stars that had fairly normal-looking spectra, but which had multiple lines as expected for magnetic white dwarfs, were put into this category.  The 107 objects in the category include some with quite unusual spectra, some that are practically featureless, and some that differ only in small ways from the DA, DB, and DO categories.  Seventy-six percent of these stars were designated as some type of white dwarf in \citet{2013yCat..22040005K}, and the others do not appear to have a designation.  We found this category of stars to be highly non-uniform, and suspect that many of these stars are magnetic - thus explaining the wide range of positions of the absorption lines.

Six spectra that looked very similar to each other were assigned the type DQ (see Harris et al. 2003, Eisenstein et al. 2006, and Kleinman et al. 2013 for sample spectra).
These stars have absorption lines of neutral carbon at 4772 \AA~and 5052 \AA.  Some also had absorption in the C I lines 4932 \AA, 5380 \AA, 6013 \AA, and 6587 \AA.  White dwarf stars with carbon absorption are called DQ WDs.  All of these stars were assigned the type DQ\_CI by \citet{2013yCat..22040005K}.

The unusual white dwarfs occupy a distinct part of the $ugr$ color-color diagram from the DA, DB, and DO WDs.  They are preferentially the fainter objects in the survey, both in apparent magnitude and reduced proper motion.  The very high proper motions of this sample of stars confirms that these stars, as a group, are closer to the Sun than any of the other classified stars.  These stars do tend to be low signal-to-noise, probably because they have faint apparent magnitudes.  However, they do not show much variability.
	
\subsection{Binary Stars}

The template matching algorithm used for SDSS DR8 fit the templates to the spectra with
the continuum subtracted.  This made it possible for very blue objects to be classified
as M dwarfs or brown dwarfs.  Typically this would happen if the signal-to-noise (S/N) was
low, or if the object was a binary star where one component was a red star.  Since the
set of templates does not contain any binary stars, it would choose either the blue or the
red star as the best match (or in one case it guessed the blue and red star together made a
galaxy).  All 129 objects in our `Binary' category have spectra indicative of a very red
star spectrum superimposed on a very blue star spectrum.  We looked for light
on both ends of the spectrum to make this classification.  With one exception of an unusual chance superposition, we did not determine whether or not
these were superpositions, or actual spectroscopic binaries.  

In the vast majority of these spectra, the blue component is a DA white dwarf, and in all of the identified binaries the red component appears to be an M star, as determined from the charateristic molecular band structure of the red side of the spectrum.  There are a few cases in which the blue star appears to be a DB white dwarf (including 2938-54526-0440, 2708-54561-0515, 2510-53877-0403, and 2467-54176-0061)\footnote{We follow the SDSS spectroscopic naming convention, where the numbers given are PLATE-MJD-FIBER; i.e., ``2938-54526-0440'' is plate 2938, MJD 54526, and fiber number 440.}.  There are also about two dozen cases where the signal-to-noise of the spectrum does not allow us to identify the blue companion.

In one case, the blue star appears to be an A star that is not a white dwarf (3148-54802-0602), paired with a brighter M star.  Because an A star - M giant star binary would be unusual, we checked the images and discovered that in this case the target was a blue star with a red star less than two arcseconds away.  This pair is likely a chance superposition rather than an actual binary star system.

Ninety-two percent of these stars were classified as a white dwarf plus M binary by \citet{2013yCat..22040005K}.  In only two cases, the 
star we classified as a binary was classified as a white dwarf (without a companion).  The proper motion distribution for binaries include a tighter distribution of more distant DA white dwarfs, and a higher proper motion set of stars that are likely closer DB and other unusual types of white dwarfs.  Given the proper motion statistics, it is likely that the stars with low signal-to-noise blue components are fainter types of dwarfs, and not DA white dwarfs.

Ten percent of the stars that are classified as binary stars show significant variability (Figures 16 and 17).  The only other categories with variability rates this high are CVs, which are also binary systems, and stars classified as DO.  We have no explanation for the high variability rate of the DO stars.

\subsection{Cataclysmic Variable}

All blue objects that had H I and/or He I emission lines but did not match the galaxy templates were categorized as CVs (compare with spectra in Shafter et al. 2011).  There are six stars identified by this criterion, plus one star (0307-51663-0090) that is listed instead in the ``Binary" category because there is faint evidence for an M star on the cool end of the spectrum.  

One of the three previously classified CVs (2603-54479-0474) has been classified as an AM CVn type binary \citep{2010PASP..122.1133S}, which is a short-period binary with a hydrogen-poor donor, and as such has prominent He I emission lines at 3889, 4471, 5876, and 6678 \AA, and no visible Balmer emission but is absent from the \citet{2011AJ....142..181S} catalog.  
Four spectra (2697-54389-0028, 2857-54453-0085, 0533-51994-0086, and 0307-51663-0090) show broad hydrogen Balmer absorption with narrow Balmer emission lines, and three of these were classified by \citet{2013yCat..22040005K} as DA WDs with emission (DAE).  Two of these stars are present in the \citet{2011AJ....142..181S} catalog, and two are not.  2857-54453-0085 is classified by Szkody et al. as ``HK Leo.''
An additional spectrum, 1039-52707-0069, was classified as a ``polar" CV by \citet{2011AJ....142..181S}.  2679-54368-0299 is present in the Szkody et al. catalog, but without a type.

The literature is unclear on the classification of systems with narrow H I and/or He I emission lines.  Some papers, like \citet{2011AJ....142..181S} (see for example SDSSJ1117 in this paper), and \citet{2013AJ....145..109H} (see blue spectrum of KIC 8490027 in this paper), classify narrow emission line objects as CVs.  \citet{2013AJ....145..109H} attribute the narrow lines to low orbital inclination, and the Balmer absorption to possibly the accretion disk or the underlying white dwarf.  Five of our seven CV stars have narrow emission lines, and of those, four show significant Balmer absorption.  If the accreting dwarf is not of type DA, then one would not expect the broad Balmer absorption, and our newly identified objects would have emission only, like most identified CVs.

On the other hand, white dwarf stars are often observed in binary systems with M stars, and the M stars can have strong, narrow emission lines, due to chromospheric activity, or due to photoionization and recombination because of irradiation from the bluer companion (see Heller et al 2009 and references therein).  \citet{2009A&A...496..191H} states that non-accreting systems do not show He I or He II features, and apparently the helium lines distinguish accreting CV stars from non-accreting binary systems with an active M star.  However, Gray \& Corbally (2009), p. 377, show a time sequence of spectra from an M9.5 dwarf flare that includes strong Balmer emission and additionally emission in many other elements including He I.

All we know for sure is that the spectra of stars in our ``Cataclysmic Variable" category, and additionally the blue part of the spectrum 0307-51663-0090, appeared similar to us and similar to other objects classified by others as type CV.  We are not certain whether any of these systems is in fact a mass transfer binary.

\subsection{B Stars}

The templates that were labeled as B6 and B9 have quite strong He I lines indicative of hotter (B0 or B3) stars.  Therefore, there was no template that was a great match to the cooler B stars.  Most of these stars were matched to one of the hot B templates, though a couple were matched to the A templates.  However, when comparing the spectra they were found not to be great matches and were therefore ``misclassified."  It is possible that some of the spectra with lower signal-to-noise (S/N$ < 15$) were also improperly matched to the hotter spectra, but would not have been flagged because the differences were too small to distinguish.  We additionally have a ``cool B" classification for stars with much stronger Balmer absorption, plus neutral helium (He I) absorption at 4026\AA, 4120\AA, 4388\AA, 4471\AA, 4713\AA, 4921\AA, and 5016\AA.  The strongest He I absorption is at 4026\AA~and 4471\AA.  Again, there is no expectation that our sample of spectra contain any young, main sequence stars; the hotter stars in the sample are exclusively from older populations and should include only white dwarf, blue straggler, horizontal branch, and other evolved stars.

The B stars have $(g-r)_0 \sim -0.5$ (the ones that are `cool' are mostly $-0.45<(g-r)_0<-0.40$), which is bluer than expected for a star of spectral type A.  Compare the colors of B stars in the Fig. 12 with the A0 stars in Fig. 6, for example.  There are $574+74=648$ stars that either match the B templates or appear to be cooler B stars in the Milky Way halo, which is interesting because these stars are all fainter than $g_0=14$ and are at high Galactic latitudes.  

Although most of the B star proper motions are consistent with no proper motion (indicative of a distant population), there are a few with measured proper motion.  Most of the stars with high proper motions are brighter than $g_0=16$, and are towards the blue side of the $(g-r)_0$ color range for these stars.  If these very blue subdwarfs are several magnitudes fainter than the turnoff, then it is possible that the brighter spectra are of stars close enough that we could measure a proper motion.  For example a star with $M_g=4$ at an apparent magnitude of $m_{g_0}=15.5$ and a tangential velocity of 100 km/s would have a proper motion of 10 mas/year.  This is telling us that the bluer of the B stars could have absolute magnitudes comparable to turnoff stars in the halo.  Alternatively, there could be a small amount of white dwarf contamination in the sample.  This is similar to the findings for stars that match the templates; the stars that match the B templates have many more proper motion outliers than the A0 and A0p stars.

\subsection{Hot Stars}

Stars classified as ``Hot'' have absorption lines of He II, in addition to fairly weak HI Balmer lines.  We separated these stars into two categories: Hot1 and Hot2.  The stars of type Hot1 have Balmer, He I, and He II lines of similar line widths, so these spectra have many absorption lines in the blue.  The He I lines at 4026\AA, 4388\AA, 4471\AA, 4921\AA, and 5886\AA~are visible, as are He II lines at 4541\AA, 4686\AA, and 5412\AA.  The stars of type Hot2 have Balmer absorption, and He II lines at 4686\AA, often 4541\AA, and sometimes 5412\AA.  Because they have He II in the spectrum, they have spectral type O, though we expect that they are subdwarf O (sdO) stars of some type.

We do not identify these stars fainter than about $g_0=18$, though in principle it should be possible.  The proper motion diagrams show that the DO stars have similar distances to the Hot (presumably sdO) stars, even though their apparent magnitudes are about the same.  Although we expected the reduced proper motion diagrams to show subdwarfs to be brighter than DO WDs, instead most of the stars are about the same, and some of them appear to be even a little fainter.  The Hot stars have similar $(g-r)_0$ colors to the DO stars, but slightly redder $(u-g)_0$ colors.  The Hot stars have very similar colors to the O stars that match the templates; in fact the O template fits pretty nicely into the set of Hot 1 spectra.  The only detectable difference we found between the stars that match the O template and the stars we have identified as Hot 1, is that the Hot 1 stars appear to be fainter in the reduced proper motion diagram.  The Hot stars do not show any sign of variability.

We have some confusion over the physical identities and evolutionary histories of the stars that fit the O templates, and those we have classified as DO, Hot 1, and Hot 2.  They must be very hot white dwarfs, very blue horizontal branch stars, post horizontal branch, or post asymptotic giant branch stars (such as the cores of planetary nebulae).  But it is unclear to us which of these stars belong in which category.

\subsection{Featureless Blue Spectra}

Spectra with a blue continuum, but that did not show significant absorption lines, were classified as ``Featureless."  A few of the stars in the category have high S/N ($>20$), and show no line features to an astounding degree.  Most have S/N $< 15$, and show hints of features - particularly hydrogen and helium, though not quite enough to make a definitive classification.  These might be classified as DC white dwarfs, which show few features.  Almost a third of these stars were classified as some type of DC white dwarf by \citet{2013yCat..22040005K}, another third were classified as some other type of white dwarf, and the rest do not appear in the white dwarf catalog.  \citet{2006ApJS..167...40E} on the other hand classified 21 of these spectra as white dwarfs of various types and 10 as subdwarf O stars.  Of the 10 stars classified as subdwarf O stars by \citet{2006ApJS..167...40E}, seven were classified as some type of white dwarf stars by \citet{2013yCat..22040005K}.  

The proper motion diagram suggests that some of the objects are brighter (and more distant), and cluster around zero proper motion.  Others have higher proper motions more like the fainter white dwarfs.  The $ugr$ color-color diagram also suggests the presence of two groups: one with colors and absolute magnitudes of DA WDs, and one that is on the very red end of our color selection.  The reduced proper motion diagram confirms that the redder stars appear to be intrinsically fainter than the bluer stars.  We confirmed by selecting stars with $(g-r)_0>-0.3$ and $(g-r)_0<-0.5$ that the redder featureless stars have the highest proper motions of any of the stellar classes seen here, while the bluer stars have proper motions that are similar to those of the fainter stars classified as hot DA white dwarfs.  This is reasonable since the featureless stars, and in particular the bluer ones, are all near the magnitude limit of the survey.

This category of stars appears to include both hot DA white dwarfs and cool DC white dwarfs, but the color-color diagram does not appear to include a substantial number of stars with colors of DB white dwarfs.  Although it is possible that a few of the stars are more similar to DO or subdwarf stars of various types, there is no evidence that there are a large number of these.  However, we agree that 2945-54505-183 does look like a PG1159 star, as it is classified by \citet{2013yCat..22040005K}.

\section{Conclusions}

We have classified all 12060 spectra of objects bluer than $(g-r)_0<-0.25$ into two catalogs: one containing objects that match one of the SDSS templates, and another grouping spectra with similar properties classified via visual examination.  The final census identifies:
\begin{itemize}
\item 294 likely sdO stars.  These include 220 stars that match the ``O" template, 56 stars labeled ``Hot 1," and 18 stars labeled ``Hot 2."
\item 648 likely sdB stars.  These include 209 stars that match the ``B6" template, 365 stars that match the ``B9" template, 66 stars labeled ``B," and 8 stars labeled ``cool B."
\item 679 likely BHB stars.  These match the ``A0p" template.
\item 1026 likely BS stars.  These match the ``A0" template.
\item 7458 DA WD stars, of which 7246 matched the ``WD" template, and 194 (primarily hot white dwarfs) are labeled ``DA," 16 are labeled ``DA+HeII," and 2 are candidate low mass DA WDs.
\item 58 magnetic DA WD stars, that match the ``WD Magnetic" template.
\item 1145 likely DB WDs, including 848 stars that match the ``OB" template and 297 stars that are labeled ``DB."
\item 92 featureless spectra, that likely include hot DA WDs and cool DC WDs.  A small number of subdwarf stars with weak absorption lines could be included in the sample.
\item 13 likely CV stars.  Seven of these match the ``CV" template and 6 are labeled ``CV."  There is an additional likely CV star that is in the ``binary" tally.
\item 129 likely WD-M dwarf binaries, labeled as ``binary."  The vast majority have a DA WD as the blue component.  One is likely a chance superposition, and not an actual binary system.  Another one is also a candidate CV star.
\item 9 likely carbon white dwarfs, which match the ``carbon WD" template.
\item 7 DZ WD stars, of various types.
\item 6 DQ WD stars.
\item 101 ``unusual WD" stars, that have non-descript or unusual broad absorption lines.  Many of these are probably magnetic white dwarfs.
\item 36 stars labeled ``DO" that seem to include a variety of hot stars, including PG1159 stars (central stars of planetary nebulae) and stars that are usually labeled DO WD.  However many of these stars appear to be intrinsically brighter than we expected for white dwarfs, so we are unsure that this classification selects a uniform set of physical objects.
\item 179 QSOs, that match one of the ``QSO" templates.
\item 10 galaxies, that match one of the emission line galaxy templates.
\item 170 low S/N objects, that we expect are primarily white dwarf stars, given the demographics of this sample.  Most of them are probably fairly featureless (without strong Balmer absorption, for example), since stars with strong absorption lines are more easily classified at low S/N.
\end{itemize}

We provide two tables of these objects, sample spectra that match the templates, figures showing all of the spectra that were grouped by eye, and diagnostic plots that show the positions, colors, apparent magnitudes, proper motions, etc. for each classification.  This catalog can be used for follow-up studies of the demographics of each population.

While most of the stars we identify as white dwarfs also appear in other white dwarf catalogs derived from the same spectra, our classifications are completely independent, and serve as a check on the other papers.  Most of the subdwarf stars are not classified in the literature.  We believe that by forcing the star to be placed in {\it some} category, we get a better sense of the numbers of each type of object.

Future surveys will be able to use templates similar to stars in each of the classes we have created to identify blue stars, including rare types, automatically.  
Additionally, the fact that blue straggler stars were apparently well separated from blue horizontal branch stars by the SDSS template-matching algorithm leads us to wonder whether this is an effective way to separate blue stars by surface gravity, which is traditionally a difficult task.

\acknowledgments

S.S. would like to thank her high school science research teacher Mrs. Regina Reals, Dr. Larry Lewis of GE Global Research, Dudley Observatory of Schenectady, NY, her science research peers and her friends and family for helping her through the task of classifying
more than twelve thousand spectra by eye.
This work was supported by the National Science Foundation, grants AST 09-37523 and AST 10-09670.  We thank the anonymous referee for helpful comments.

Funding for SDSS-III has been provided by the Alfred P. Sloan Foundation, the Participating Institutions, the National Science Foundation, and the U.S. Department of Energy Office of Science. The SDSS-III web site is http://www.sdss3.org/.
SDSS-III is managed by the Astrophysical Research Consortium for the Participating Institutions of the SDSS-III Collaboration including the University of Arizona, the Brazilian Participation Group, Brookhaven National Laboratory, University of Cambridge, Carnegie Mellon University, University of Florida, the French Participation Group, the German Participation Group, Harvard University, the Instituto de Astrofisica de Canarias, the Michigan State/Notre Dame/JINA Participation Group, Johns Hopkins University, Lawrence Berkeley National Laboratory, Max Planck Institute for Astrophysics, Max Planck Institute for Extraterrestrial Physics, New Mexico State University, New York University, Ohio State University, Pennsylvania State University, University of Portsmouth, Princeton University, the Spanish Participation Group, University of Tokyo, University of Utah, Vanderbilt University, University of Virginia, University of Washington, and Yale University.

\newpage

\newpage

\newpage


\setcounter{figure}{2}
\begin{figure}
\includegraphics[width=0.95\textwidth]{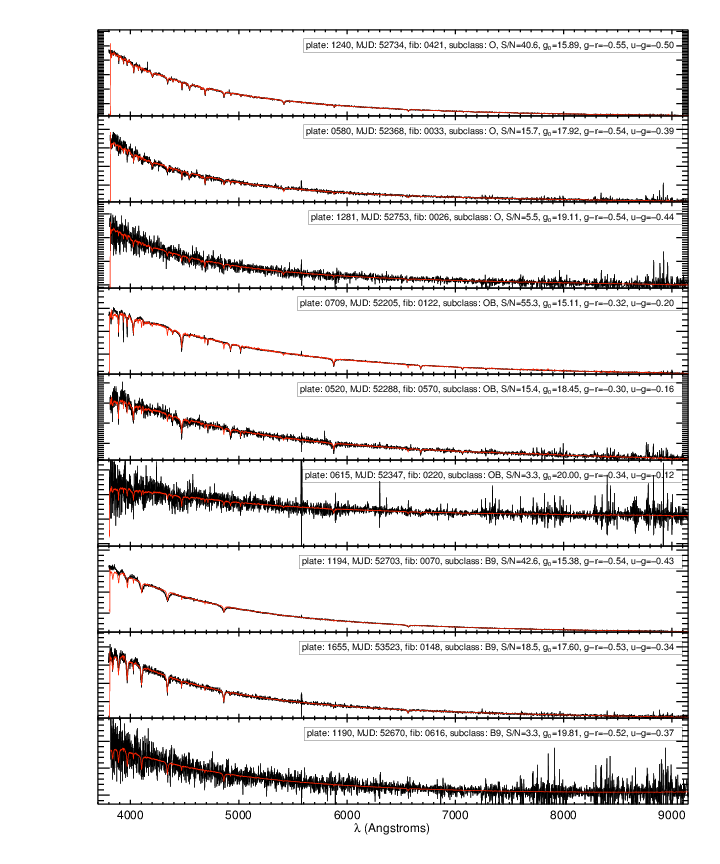}
\caption{(a) Sample spectra from each category of stars that matched the SDSS templates. Each panel shows an SDSS spectrum in black, with the matching template overlaid as a red line. The plate, MJD, and fiber, as well as the magnitude and colors, are given for each object. The object class is identified by the ``subclass'' label in the legend of each panel. For each class, we show spectra spanning a range of S/N (except for CVs and carbon WDs, all of which are shown).}
\end{figure}


\setcounter{figure}{2}
\begin{figure}
\includegraphics[width=0.95\textwidth]{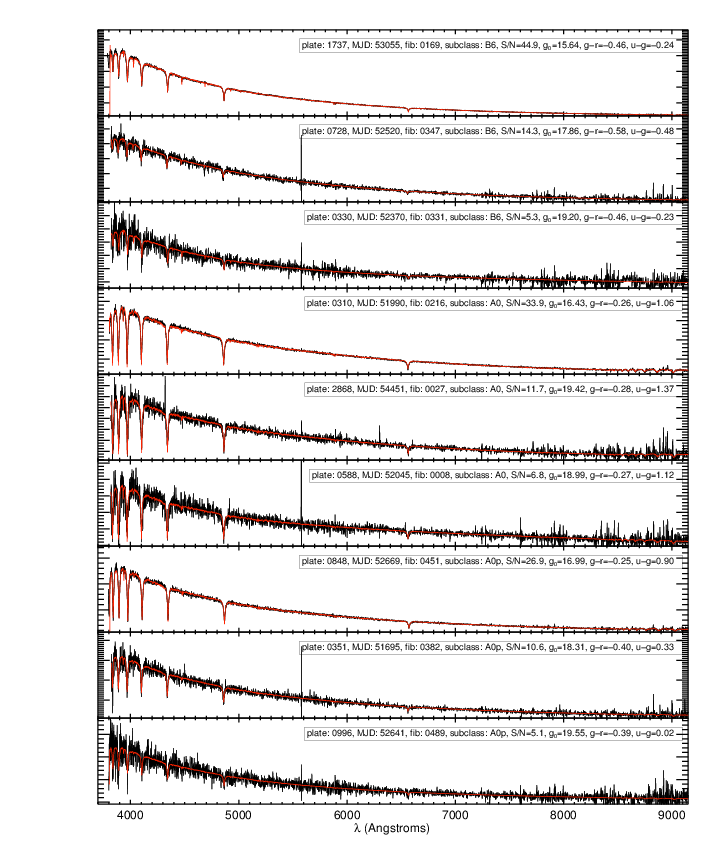}
\caption{(b) Sample template-matched spectra, continued...}
\end{figure}


\setcounter{figure}{2}
\begin{figure}
\includegraphics[width=0.95\textwidth]{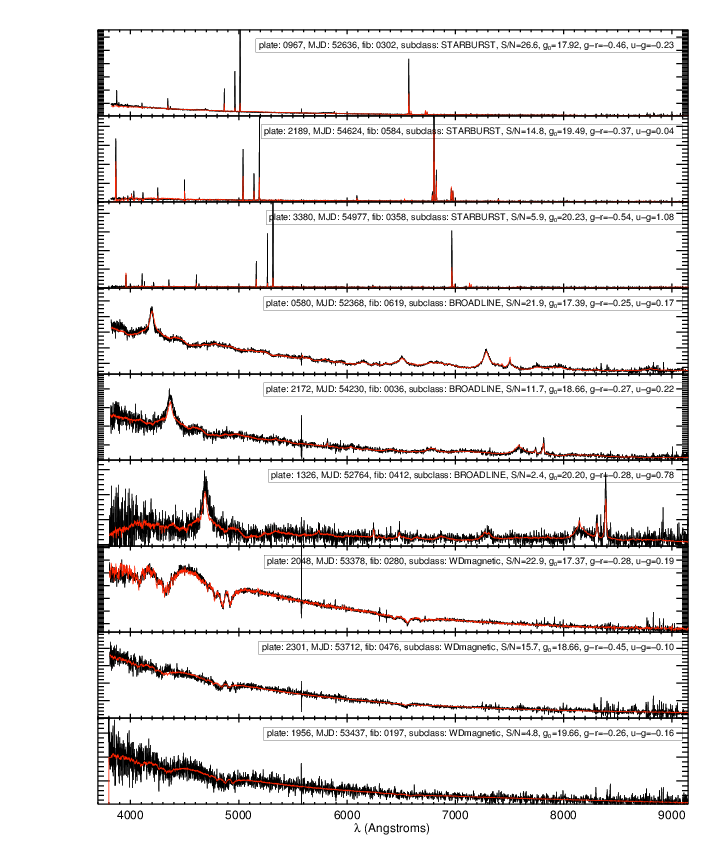}
\caption{(c) Sample template-matched spectra, continued...}
\end{figure}


\setcounter{figure}{2}
\begin{figure}
\includegraphics[width=0.95\textwidth]{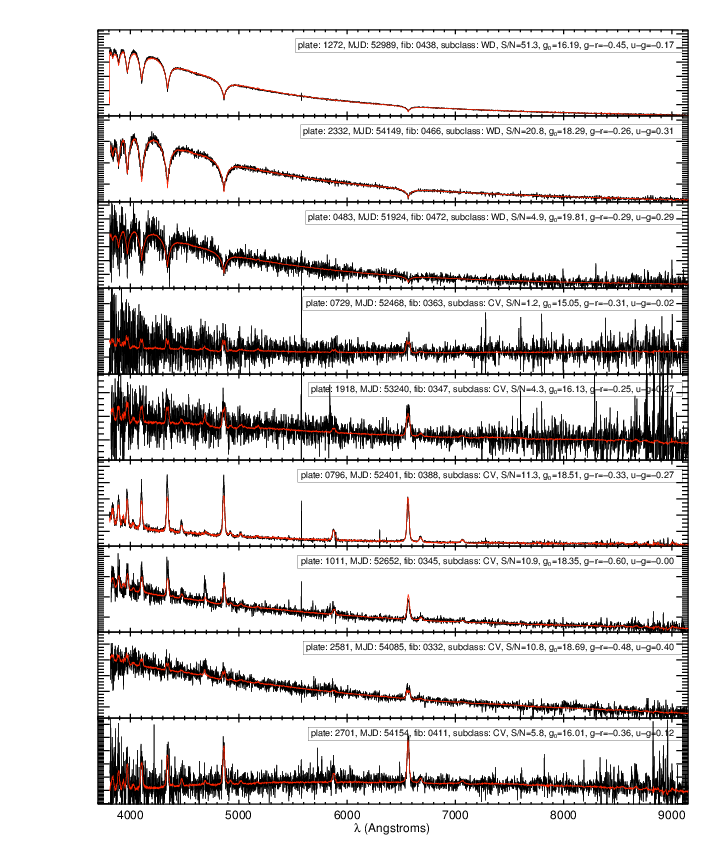}
\caption{(d) Sample template-matched spectra, continued...}
\end{figure}


\setcounter{figure}{2}
\begin{figure}
\includegraphics[width=0.95\textwidth]{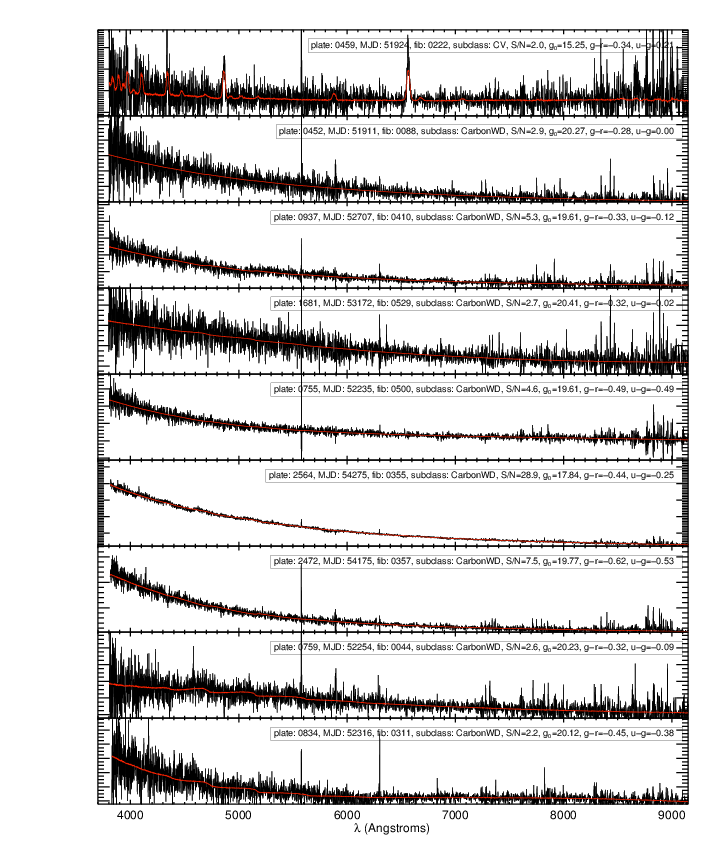}
\caption{(e) Sample template-matched spectra, continued...}
\end{figure}


\setcounter{figure}{2}
\begin{figure}
\includegraphics[width=0.95\textwidth]{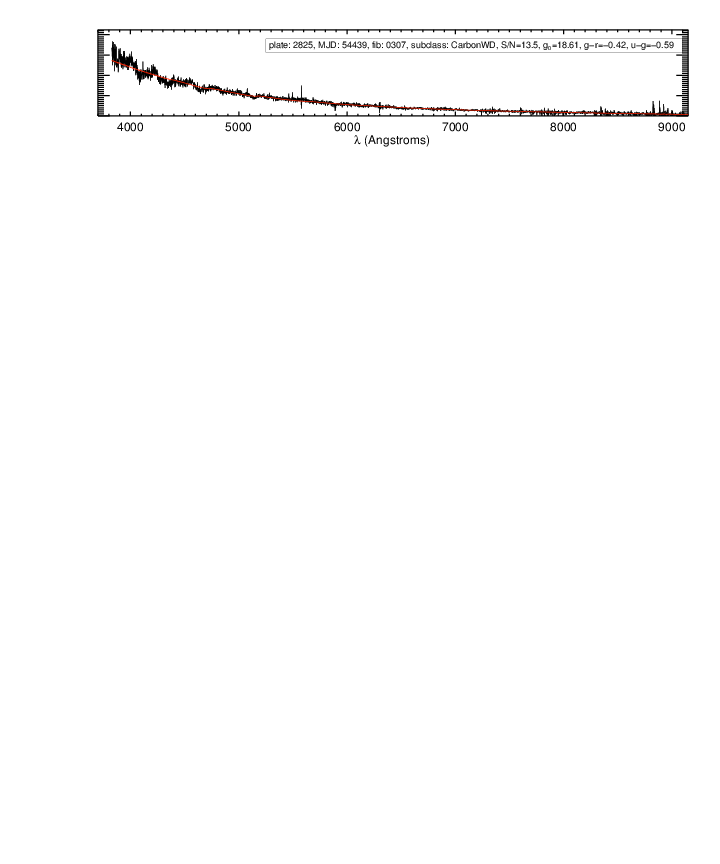}
\caption{(f) Sample template-matched spectra, continued...}
\end{figure}

\newpage


\setcounter{figure}{3}
\begin{figure}
\includegraphics[width=0.95\textwidth]{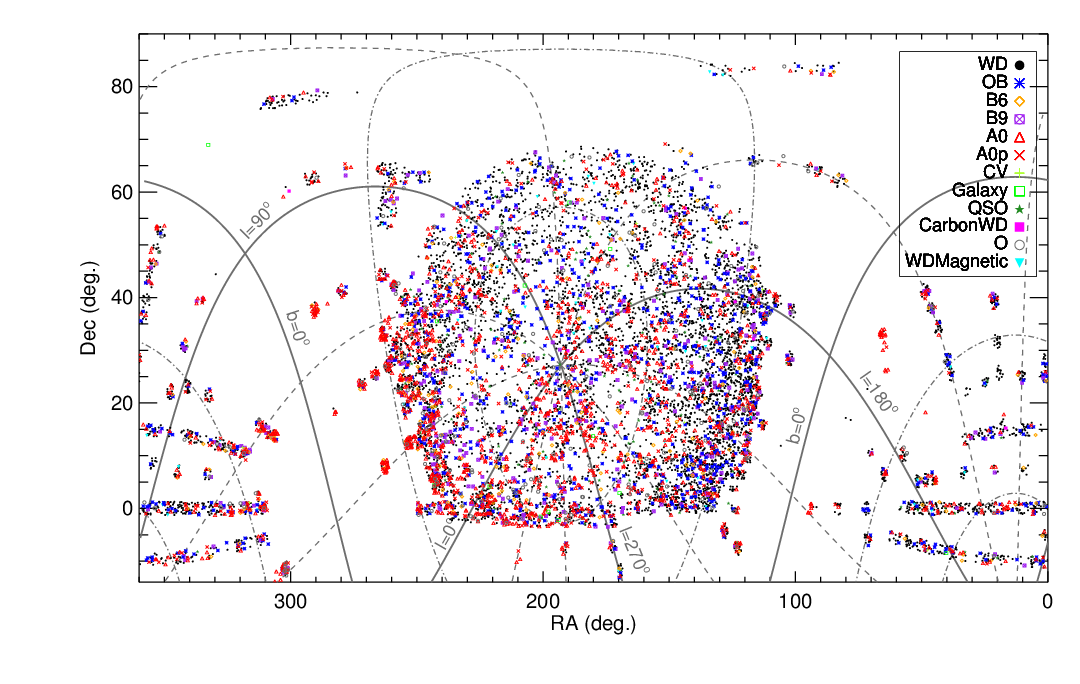}
\caption{Positions (in equatorial coordinates) of objects that correctly matched the SDSS templates (from Table 4), separated by classification (see the legend for symbol types assigned to different classes).  Galactic coordinates are shown for reference.}
\end{figure}

\newpage


\begin{figure}
\includegraphics[width=5.0in]{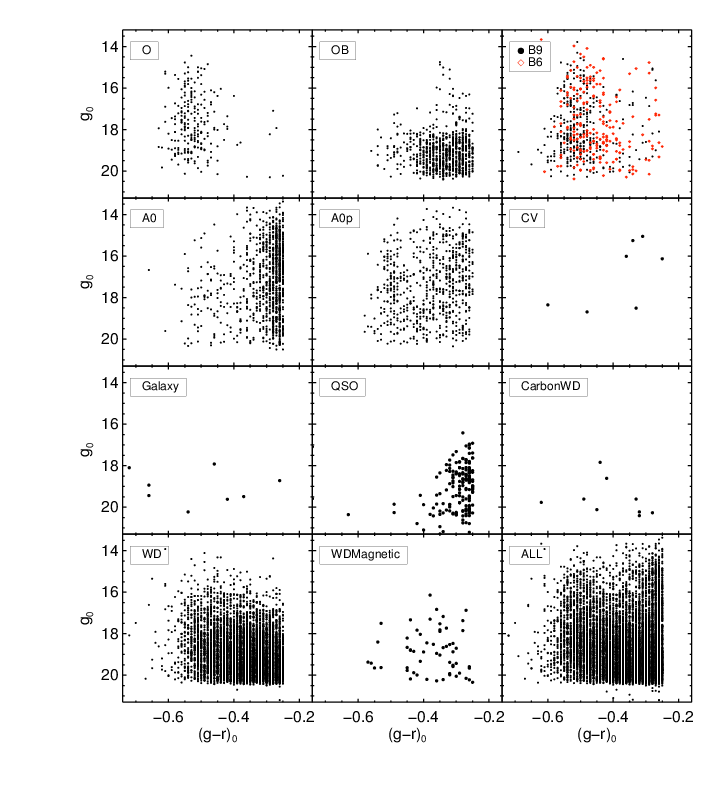}
\caption{Color magnitude diagrams of objects that correctly matched the SDSS templates (from Table 4), separated by classification.}
\end{figure}

\newpage


\begin{figure}[!ht]
\includegraphics[width=5.0in]{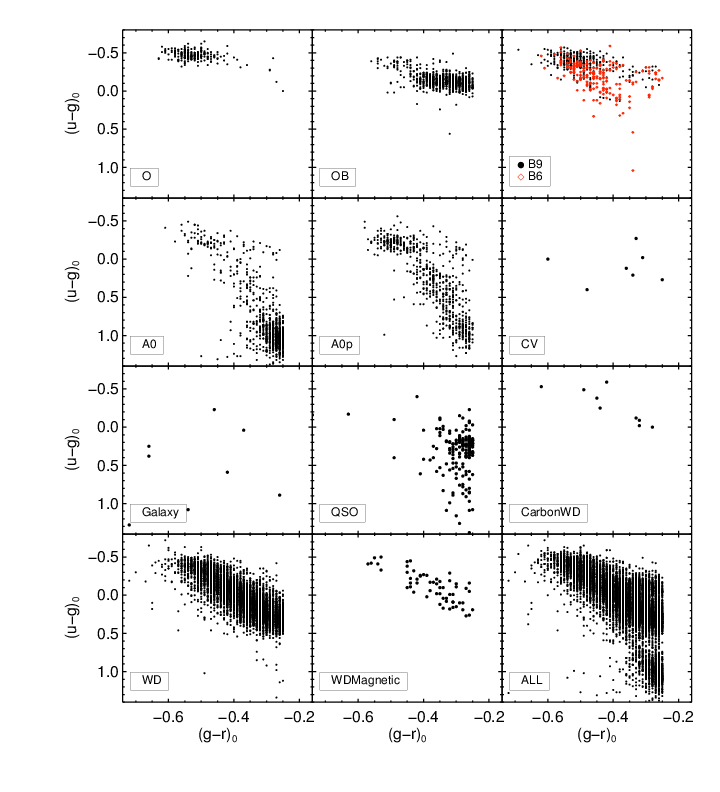}
\caption{Color-color diagrams of objects that correctly matched the SDSS templates (from Table 4), separated by classification.}
\end{figure}

\newpage


\begin{figure}[!ht]
\includegraphics[width=5.0in]{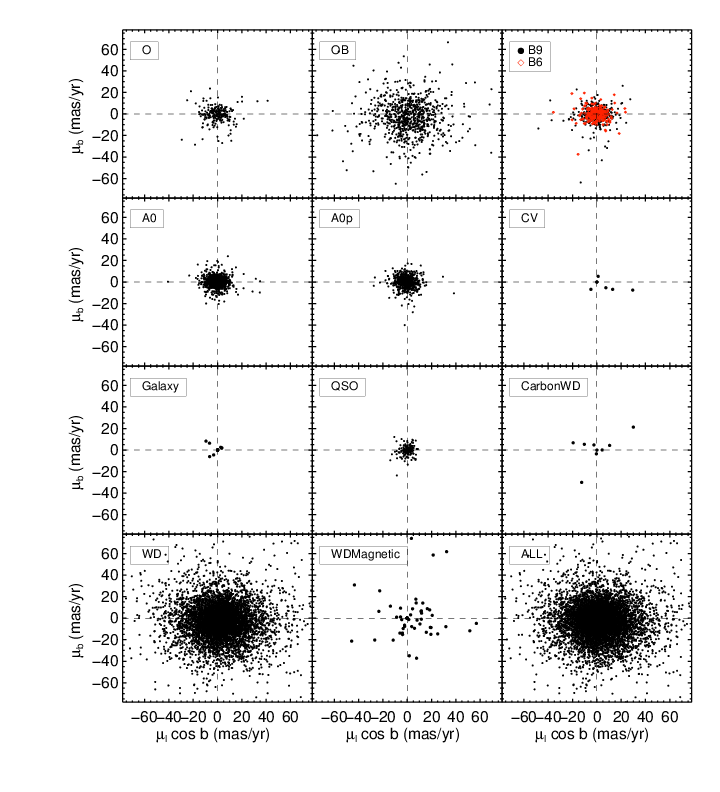}
\caption{Proper motion vector point diagrams of objects that correctly matched the SDSS templates (from Table 4), separated by classification.}
\end{figure}

\newpage


\begin{figure}[!ht]
\includegraphics[width=5.0in]{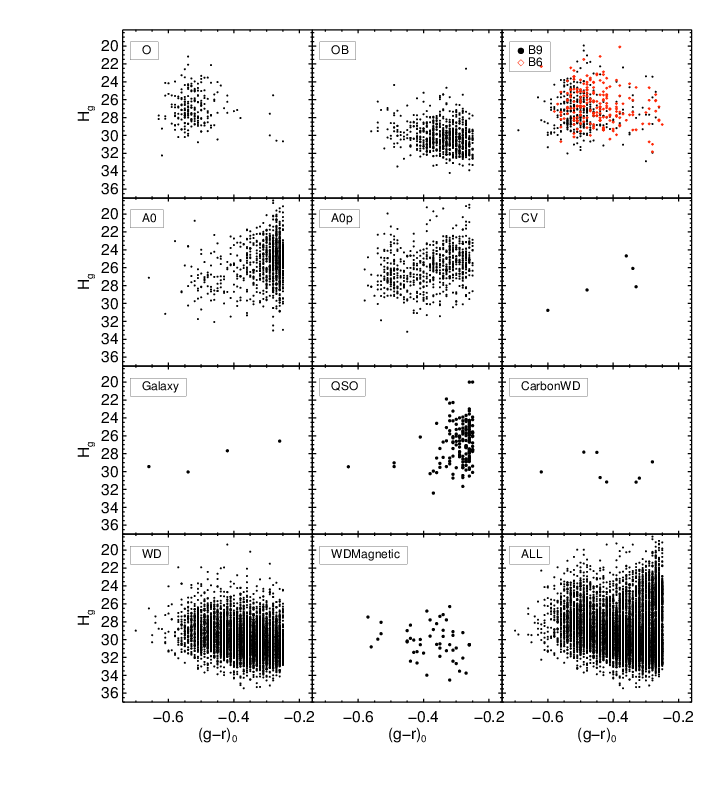}
\caption{Reduced proper motion diagrams of objects that correctly matched the SDSS templates (from Table 4), separated by classification.}
\end{figure}

\newpage


\begin{figure}[!ht]
\includegraphics[width=5.0in]{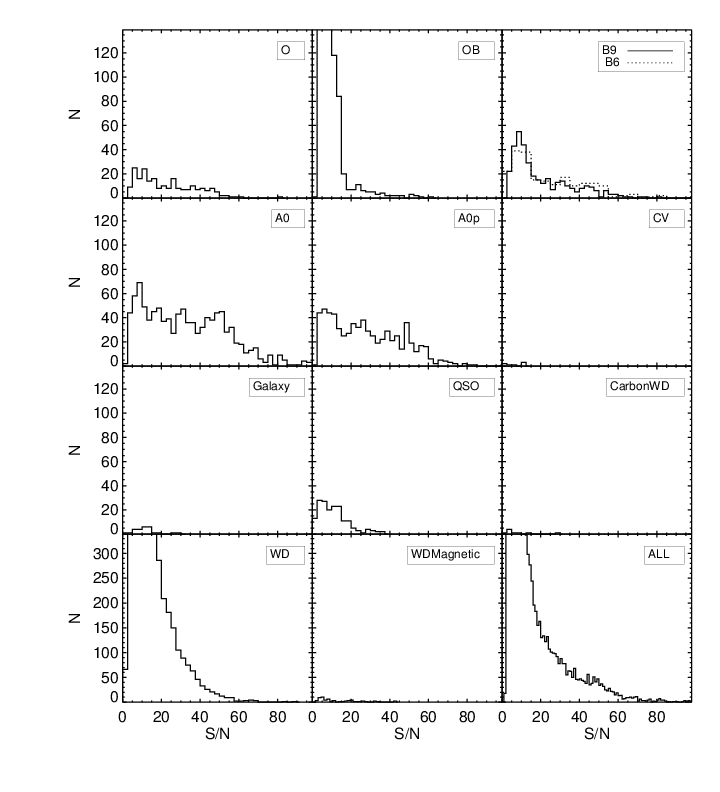}
\caption{Signal to noise of objects that correctly matched the SDSS templates (from Table 4), separated by classification.}
\end{figure}

\newpage


\begin{figure}[!ht]
\includegraphics[width=0.95\textwidth]{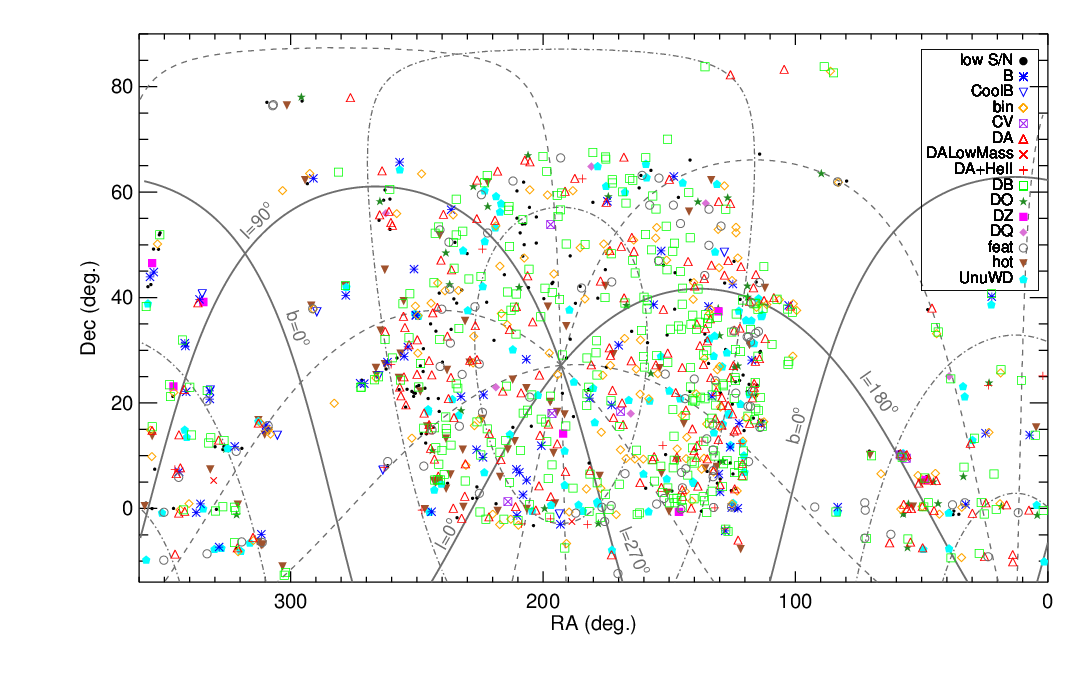}
\caption{Positions (in equatorial coordinates) of stars that were misclassified by the SDSS template matching (from Table 5), separated by the new classification we have assigned to them (see the legend for symbol types).  Galactic coordinates are shown for reference.}
\end{figure}


\begin{figure}[!ht]
\includegraphics[width=5.0in]{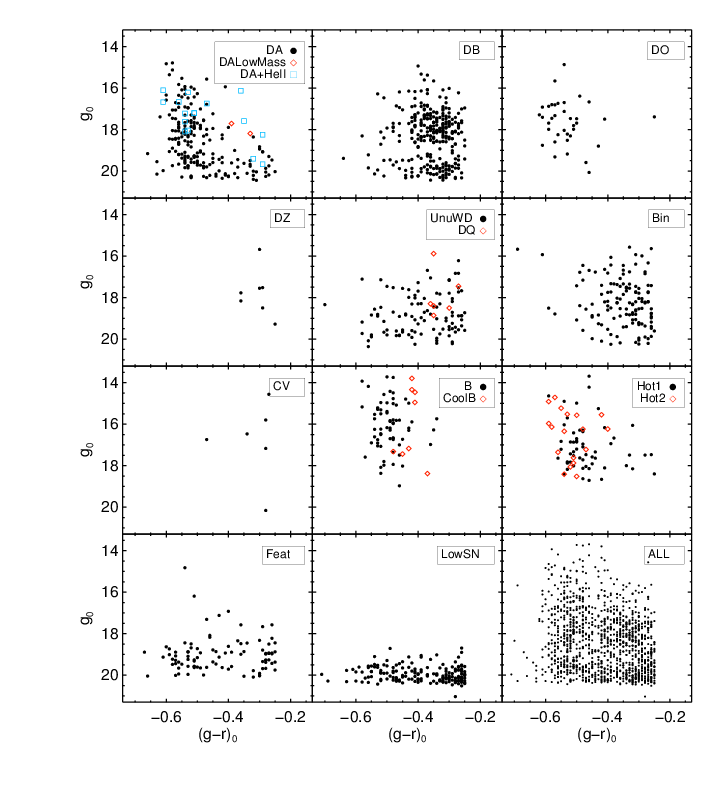}
\caption{Color magnitude diagrams of reclassified stars, separated by their new classification.}
\end{figure}


\begin{figure}[!ht]
\includegraphics[width=5.0in]{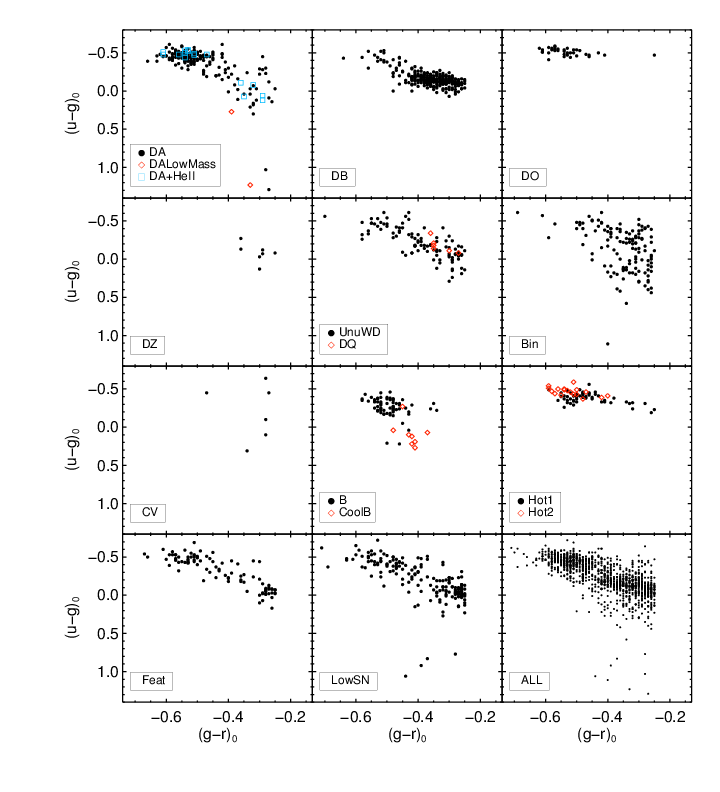}
\caption{Color-color diagrams of reclassified stars, separated by their new classification.}
\end{figure}


\begin{figure}[!ht]
\includegraphics[width=5.0in]{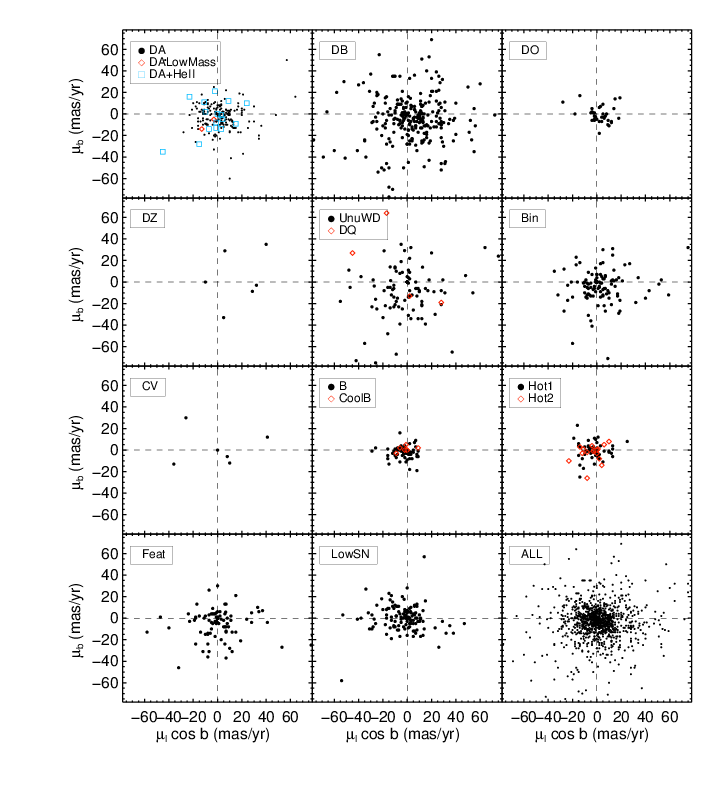}
\caption{Proper motion vector point diagrams of reclassified stars, separated by their new classification.}
\end{figure}

\clearpage


\newpage
\begin{figure}[!ht]
\includegraphics[width=5.0in]{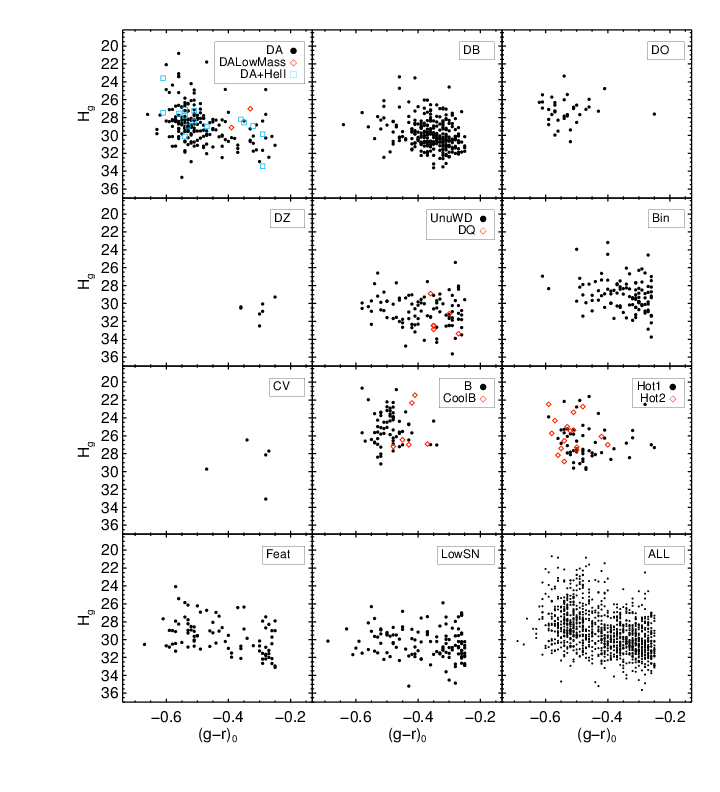}
\caption{Reduced proper motion diagrams of reclassified stars, separated by their new classification.}
\end{figure}

\newpage
\begin{figure}[!ht]
\includegraphics[width=5.0in]{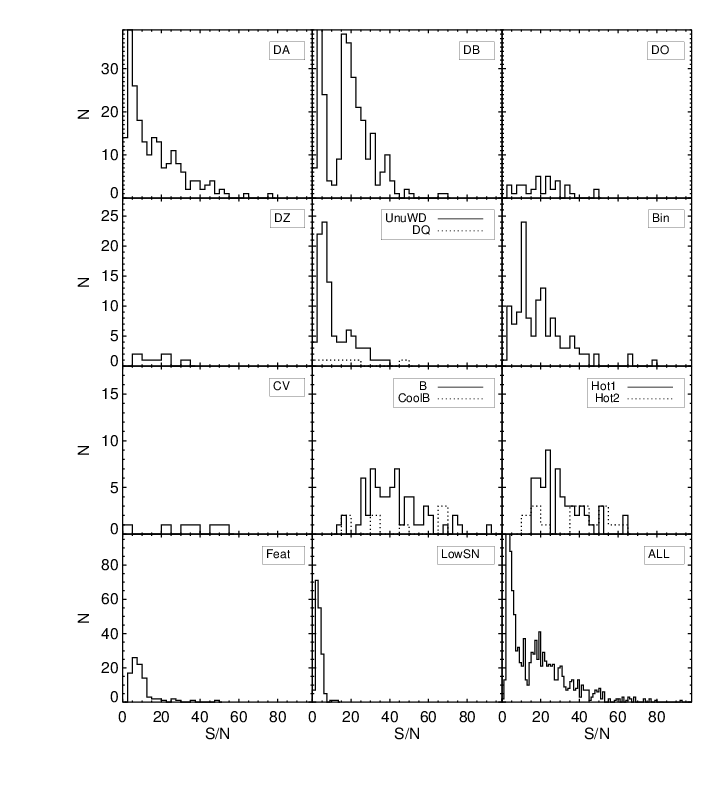}
\caption{Signal to noise of reclassified stars, separated by their new classification.}
\end{figure}

\newpage
\begin{figure}[!ht]
\includegraphics[width=5.0in]{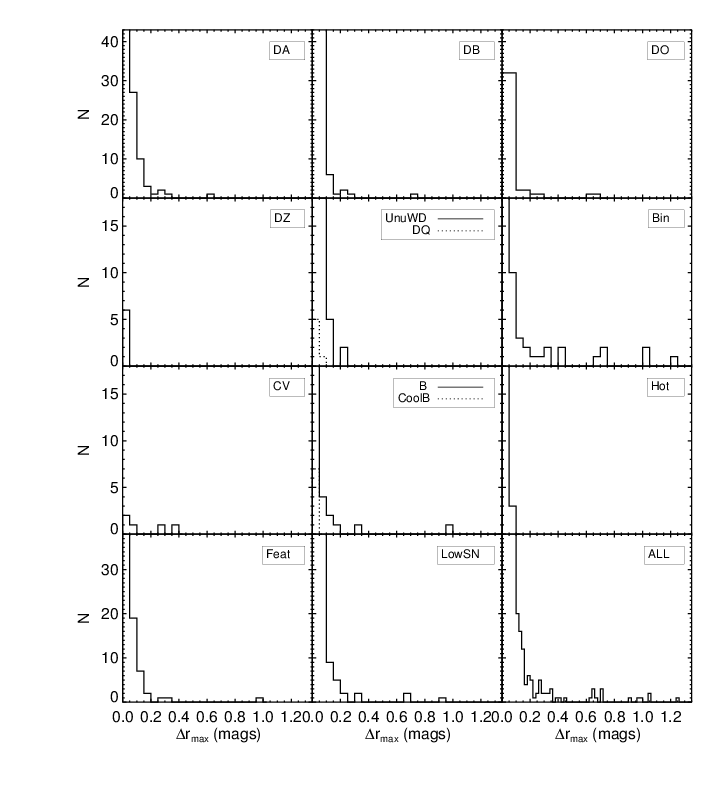}
\caption{Maximum $r$-magnitude difference $\Delta$r$_{\rm max}$ of all measurements for the reclassified stars, separated by their new classification.}
\end{figure}


\newpage
\begin{figure}[!ht]
\includegraphics[width=5.0in]{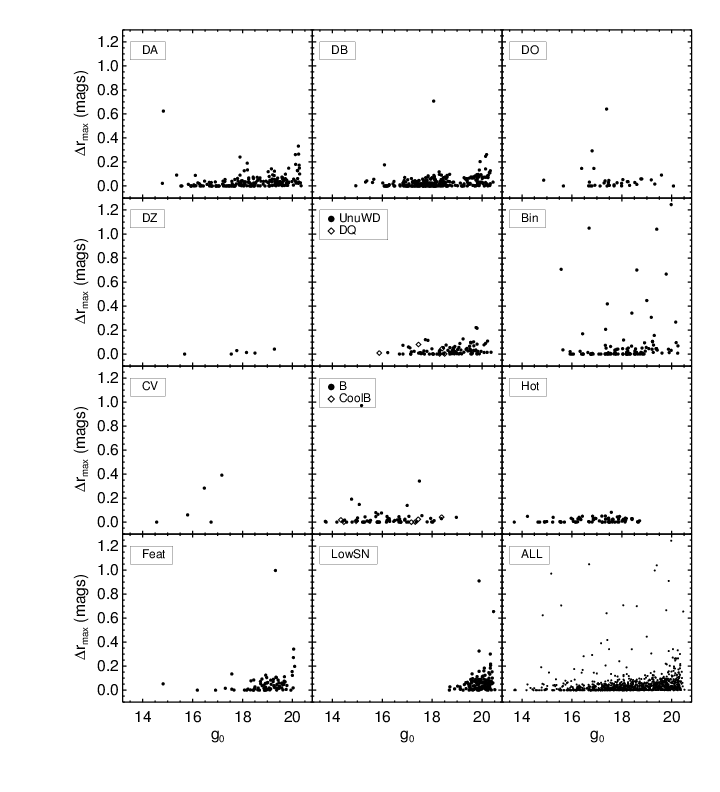}
\caption{Maximum $r$-magnitude difference $\Delta$r$_{\rm max}$ as a function of $g$ magnitude for all reclassified stars, separated by classification.}
\end{figure}


\newpage

\figsetstart
\figsetnum{18}
\figsettitle{Spectra of stars classified as ``LowSN''}

\figsetgrpstart
\figsetgrpnum{18.1}
\figsetgrptitle{Page 1 of class ``LowSN'' spectra}
\figsetplot{LowSN_spectra0_jun2014.png}
\figsetgrpnote{Spectra of all stars classified as ``LowSN'' in Table 5.}
\figsetgrpend

\figsetgrpstart
\figsetgrpnum{18.2}
\figsetgrptitle{Page 2 of class ``LowSN'' spectra}
\figsetplot{LowSN_spectra1_jun2014.png}
\figsetgrpnote{Spectra of all stars classified as ``LowSN'' in Table 5.}
\figsetgrpend

\figsetgrpstart
\figsetgrpnum{18.3}
\figsetgrptitle{Page 3 of class ``LowSN'' spectra}
\figsetplot{LowSN_spectra2_jun2014.png}
\figsetgrpnote{Spectra of all stars classified as ``LowSN'' in Table 5.}
\figsetgrpend

\figsetgrpstart
\figsetgrpnum{18.4}
\figsetgrptitle{Page 4 of class ``LowSN'' spectra}
\figsetplot{LowSN_spectra3_jun2014.png}
\figsetgrpnote{Spectra of all stars classified as ``LowSN'' in Table 5.}
\figsetgrpend

\figsetgrpstart
\figsetgrpnum{18.5}
\figsetgrptitle{Page 5 of class ``LowSN'' spectra}
\figsetplot{LowSN_spectra4_jun2014.png}
\figsetgrpnote{Spectra of all stars classified as ``LowSN'' in Table 5.}
\figsetgrpend

\figsetend

\begin{figure}[!ht]
\figurenum{18}
\includegraphics[width=5.0in]{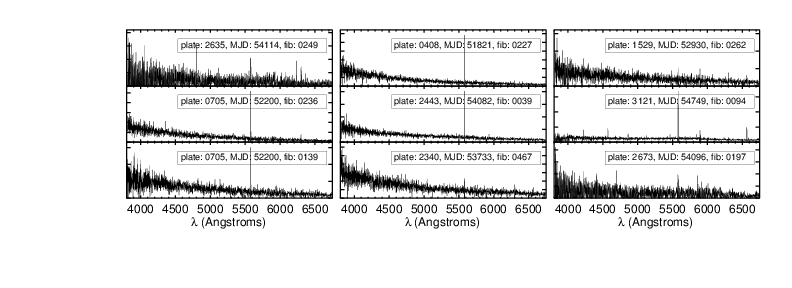}
\caption{Spectra of stars classified as ``LowSN'' in Table 5. Figures 18.1 - 18.5, showing the entire set of spectra, are available in the online version of the Journal.}
\end{figure}

\clearpage
\newpage

\figsetstart
\figsetnum{19}
\figsettitle{Spectra of stars classified as ``DA''}

\figsetgrpstart
\figsetgrpnum{19.1}
\figsetgrptitle{Page 1 of class ``DA'' spectra}
\figsetplot{DA_spectra0_jun2014.png}
\figsetgrpnote{Spectra of all stars classified as ``DA'' in Table 5.}
\figsetgrpend

\figsetgrpstart
\figsetgrpnum{19.2}
\figsetgrptitle{Page 2 of class ``DA'' spectra}
\figsetplot{DA_spectra1_jun2014.png}
\figsetgrpnote{Spectra of all stars classified as ``DA'' in Table 5.}
\figsetgrpend

\figsetgrpstart
\figsetgrpnum{19.3}
\figsetgrptitle{Page 3 of class ``DA'' spectra}
\figsetplot{DA_spectra2_jun2014.png}
\figsetgrpnote{Spectra of all stars classified as ``DA'' in Table 5.}
\figsetgrpend

\figsetgrpstart
\figsetgrpnum{19.4}
\figsetgrptitle{Page 4 of class ``DA'' spectra}
\figsetplot{DA_spectra3_jun2014.png}
\figsetgrpnote{Spectra of all stars classified as ``DA'' in Table 5.}
\figsetgrpend

\figsetgrpstart
\figsetgrpnum{19.5}
\figsetgrptitle{Page 5 of class ``DA'' spectra}
\figsetplot{DA_spectra4_jun2014.png}
\figsetgrpnote{Spectra of all stars classified as ``DA'' in Table 5.}
\figsetgrpend

\figsetend

\begin{figure}[!ht]
\figurenum{19}
\includegraphics[width=5.0in]{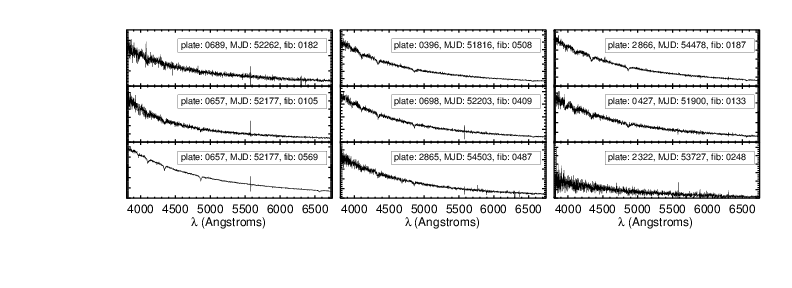}
\caption{Spectra of stars classified as ``DA'' in Table 5. Figures 19.1 - 19.5, showing the entire set of spectra, are available in the online version of the Journal.}
\end{figure}

\newpage

\figsetstart
\figsetnum{20}
\figsettitle{Spectra of stars classified as ``DB''}

\figsetgrpstart
\figsetgrpnum{20.1}
\figsetgrptitle{Page 1 of class ``DB'' spectra}
\figsetplot{DB_spectra0_jun2014.png}
\figsetgrpnote{Spectra of all stars classified as ``DB'' in Table 5.}
\figsetgrpend

\figsetgrpstart
\figsetgrpnum{20.2}
\figsetgrptitle{Page 2 of class ``DB'' spectra}
\figsetplot{DB_spectra1_jun2014.png}
\figsetgrpnote{Spectra of all stars classified as ``DB'' in Table 5.}
\figsetgrpend

\figsetgrpstart
\figsetgrpnum{20.3}
\figsetgrptitle{Page 3 of class ``DB'' spectra}
\figsetplot{DB_spectra2_jun2014.png}
\figsetgrpnote{Spectra of all stars classified as ``DB'' in Table 5.}
\figsetgrpend

\figsetgrpstart
\figsetgrpnum{20.4}
\figsetgrptitle{Page 4 of class ``DB'' spectra}
\figsetplot{DB_spectra3_jun2014.png}
\figsetgrpnote{Spectra of all stars classified as ``DB'' in Table 5.}
\figsetgrpend

\figsetgrpstart
\figsetgrpnum{20.5}
\figsetgrptitle{Page 5 of class ``DB'' spectra}
\figsetplot{DB_spectra4_jun2014.png}
\figsetgrpnote{Spectra of all stars classified as ``DB'' in Table 5.}
\figsetgrpend

\figsetgrpstart
\figsetgrpnum{20.6}
\figsetgrptitle{Page 6 of class ``DB'' spectra}
\figsetplot{DB_spectra5_jun2014.png}
\figsetgrpnote{Spectra of all stars classified as ``DB'' in Table 5.}
\figsetgrpend

\figsetgrpstart
\figsetgrpnum{20.7}
\figsetgrptitle{Page 7 of class ``DB'' spectra}
\figsetplot{DB_spectra6_jun2014.png}
\figsetgrpnote{Spectra of all stars classified as ``DB'' in Table 5.}
\figsetgrpend

\figsetend

\begin{figure}[!ht]
\figurenum{20}
\includegraphics[width=5.0in]{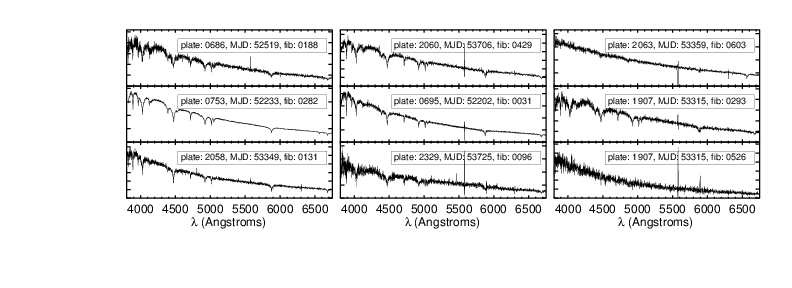}
\caption{Spectra of stars classified as ``DB'' in Table 5. Figures 20.1 - 20.7, showing the entire set of spectra, are available in the online version of the Journal.}
\end{figure}

\newpage
\begin{figure}[!ht]
\figurenum{21}
\includegraphics[width=5.0in]{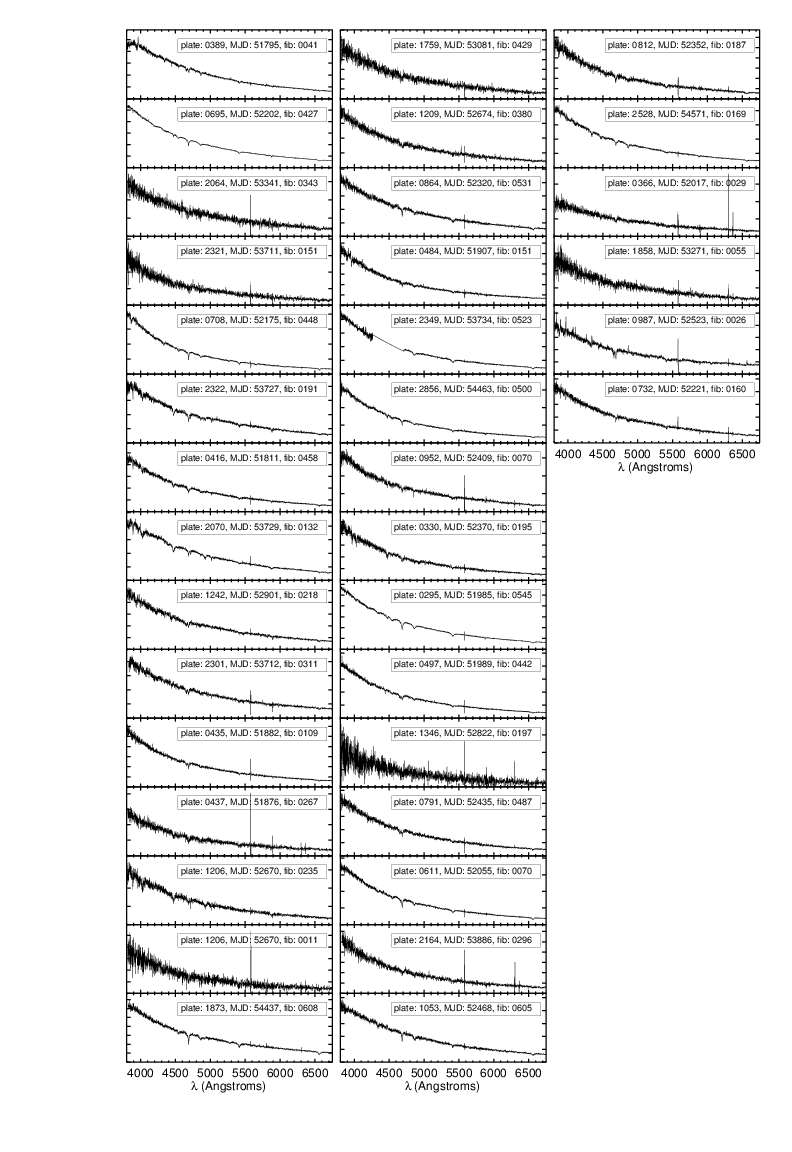}
\caption{Spectra of all stars classified as ``DO'' in Table 5.}
\end{figure}

\newpage
\begin{figure}[!ht]
\figurenum{22}
\includegraphics[width=5.0in]{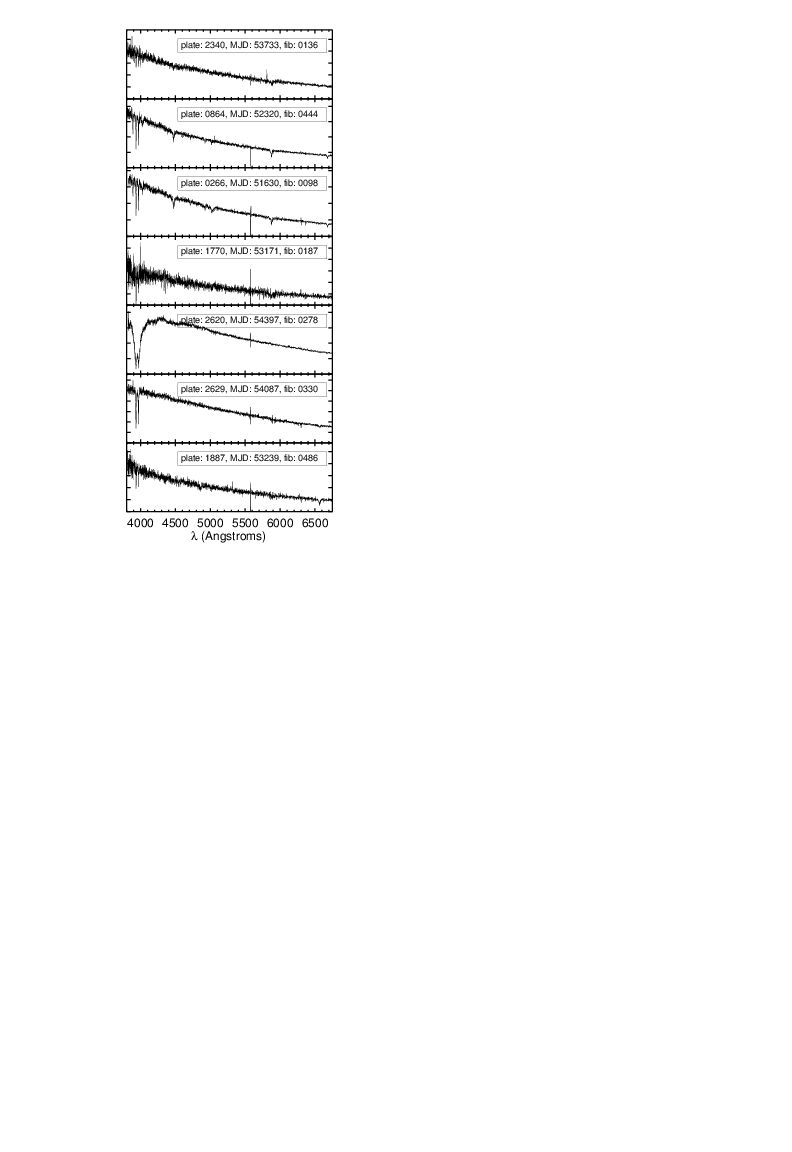}
\caption{Spectra of all stars classified as ``DZ'' in Table 5.}
\end{figure}


\newpage
\figsetstart
\figsetnum{23}
\figsettitle{Spectra of stars classified as ``UnuWD''}

\figsetgrpstart
\figsetgrpnum{23.1}
\figsetgrptitle{Page 1 of class ``UnuWD'' spectra}
\figsetplot{UnuWD_spectra0_jun2014.png}
\figsetgrpnote{Spectra of all stars classified as ``UnuWD'' in Table 5.}
\figsetgrpend

\figsetgrpstart
\figsetgrpnum{23.2}
\figsetgrptitle{Page 2 of class ``UnuWD'' spectra}
\figsetplot{UnuWD_spectra1_jun2014.png}
\figsetgrpnote{Spectra of all stars classified as ``UnuWD'' in Table 5.}
\figsetgrpend

\figsetgrpstart
\figsetgrpnum{23.3}
\figsetgrptitle{Page 3 of class ``UnuWD'' spectra}
\figsetplot{UnuWD_spectra2_jun2014.png}
\figsetgrpnote{Spectra of all stars classified as ``UnuWD'' in Table 5.}
\figsetgrpend

\figsetend

\begin{figure}[!ht]
\figurenum{23}
\includegraphics[width=5.0in]{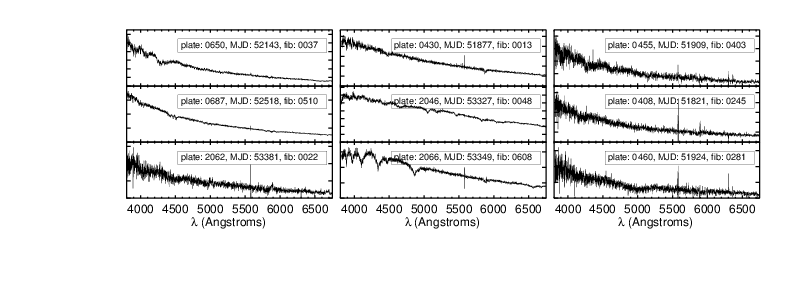}
\caption{Spectra of stars classified as ``UnuWD'' in Table 5. Figures 23.1 - 23.3, showing the entire set of spectra, are available in the online version of the Journal.}
\end{figure}

\newpage
\figsetstart
\figsetnum{24}
\figsettitle{Spectra of stars classified as ``Bin''}

\figsetgrpstart
\figsetgrpnum{24.1}
\figsetgrptitle{Page 1 of class ``Bin'' spectra}
\figsetplot{Bin_spectra0_jun2014.png}
\figsetgrpnote{Spectra of all stars classified as ``Bin'' in Table 5.}
\figsetgrpend

\figsetgrpstart
\figsetgrpnum{24.2}
\figsetgrptitle{Page 2 of class ``Bin'' spectra}
\figsetplot{Bin_spectra1_jun2014.png}
\figsetgrpnote{Spectra of all stars classified as ``Bin'' in Table 5.}
\figsetgrpend

\figsetgrpstart
\figsetgrpnum{24.3}
\figsetgrptitle{Page 3 of class ``Bin'' spectra}
\figsetplot{Bin_spectra2_jun2014.png}
\figsetgrpnote{Spectra of all stars classified as ``Bin'' in Table 5.}
\figsetgrpend

\figsetend

\begin{figure}[!ht]
\figurenum{24}
\includegraphics[width=5.0in]{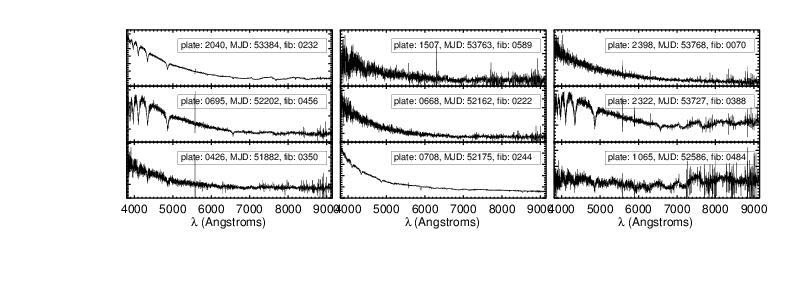}
\caption{Spectra of stars classified as ``Bin'' in Table 5. Figures 24.1 - 24.3, showing the entire set of spectra, are available in the online version of the Journal.}
\end{figure}

\begin{figure}[!ht]
\figurenum{25}
\includegraphics[width=5.0in]{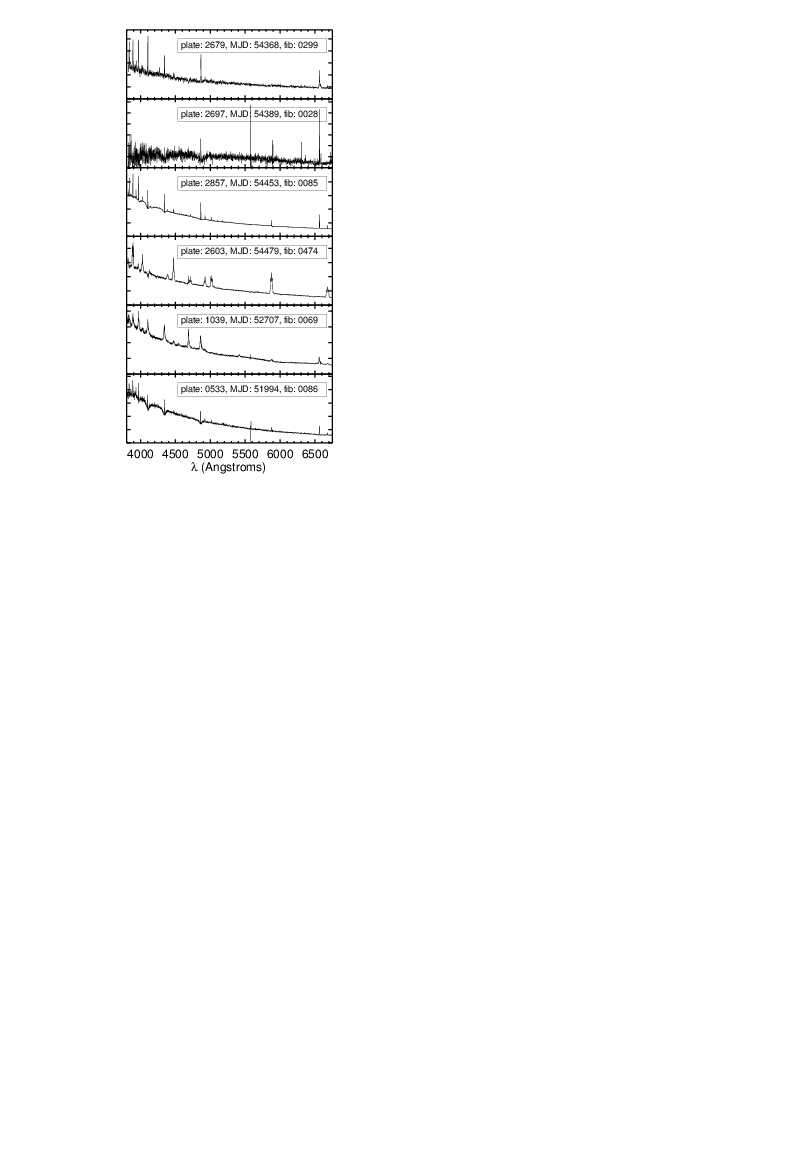}
\caption{Spectra of all stars classified as ``CV'' in Table 5.}
\end{figure}

\clearpage
\newpage
\figsetstart
\figsetnum{26}
\figsettitle{Spectra of stars classified as ``B''}

\figsetgrpstart
\figsetgrpnum{26.1}
\figsetgrptitle{Page 1 of class ``B'' spectra}
\figsetplot{B_spectra0_jun2014.png}
\figsetgrpnote{Spectra of all stars classified as ``B'' in Table 5.}
\figsetgrpend

\figsetgrpstart
\figsetgrpnum{26.2}
\figsetgrptitle{Page 2 of class ``B'' spectra}
\figsetplot{B_spectra1_jun2014.png}
\figsetgrpnote{Spectra of all stars classified as ``B'' in Table 5.}
\figsetgrpend

\figsetend

\begin{figure}[!ht]
\figurenum{26}
\includegraphics[width=5.0in]{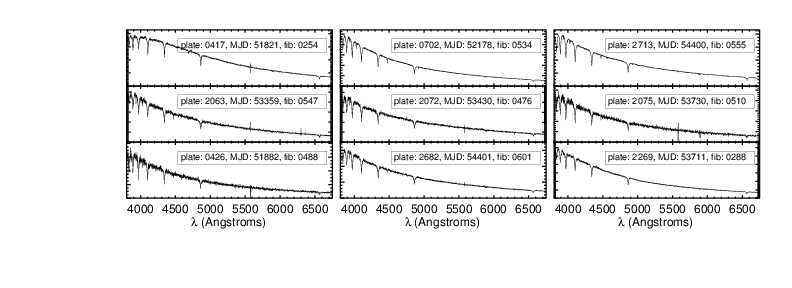}
\caption{Spectra of stars classified as ``B'' in Table 5. Figures 26.1 - 26.2, showing the entire set of spectra, are available in the online version of the Journal.}
\end{figure}

\newpage
\figsetstart
\figsetnum{27}
\figsettitle{Spectra of stars classified as ``Hot''}

\figsetgrpstart
\figsetgrpnum{27.1}
\figsetgrptitle{Page 1 of class ``Hot'' spectra}
\figsetplot{Hot_spectra0_jun2014.png}
\figsetgrpnote{Spectra of all stars classified as ``Hot'' in Table 5.}
\figsetgrpend

\figsetgrpstart
\figsetgrpnum{27.2}
\figsetgrptitle{Page 2 of class ``Hot'' spectra}
\figsetplot{Hot_spectra1_jun2014.png}
\figsetgrpnote{Spectra of all stars classified as ``Hot'' in Table 5.}
\figsetgrpend

\figsetend

\begin{figure}[!ht]
\figurenum{27}
\includegraphics[width=5.0in]{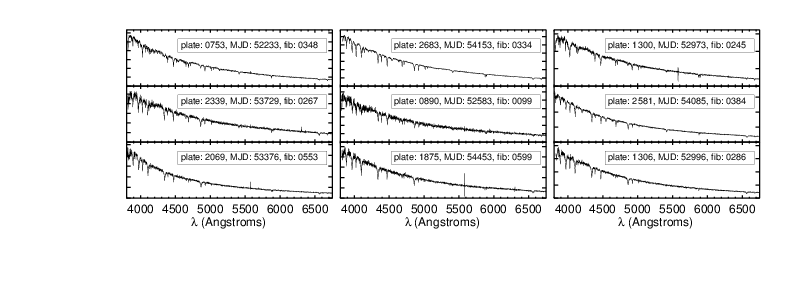}
\caption{Spectra of stars classified as ``Hot'' in Table 5. Figures 27.1 - 27.2, showing the entire set of spectra, are available in the online version of the Journal.}
\end{figure}

\newpage
\figsetstart
\figsetnum{28}
\figsettitle{Spectra of stars classified as ``Feat''}

\figsetgrpstart
\figsetgrpnum{28.1}
\figsetgrptitle{Page 1 of class ``Feat'' spectra}
\figsetplot{Feat_spectra0_jun2014.png}
\figsetgrpnote{Spectra of all stars classified as ``Feat'' in Table 5.}
\figsetgrpend

\figsetgrpstart
\figsetgrpnum{28.2}
\figsetgrptitle{Page 2 of class ``Feat'' spectra}
\figsetplot{Feat_spectra1_jun2014.png}
\figsetgrpnote{Spectra of all stars classified as ``Feat'' in Table 5.}
\figsetgrpend

\figsetgrpstart
\figsetgrpnum{28.3}
\figsetgrptitle{Page 3 of class ``Feat'' spectra}
\figsetplot{Feat_spectra2_jun2014.png}
\figsetgrpnote{Spectra of all stars classified as ``Feat'' in Table 5.}
\figsetgrpend

\figsetend

\begin{figure}[!ht]
\figurenum{28}
\includegraphics[width=5.0in]{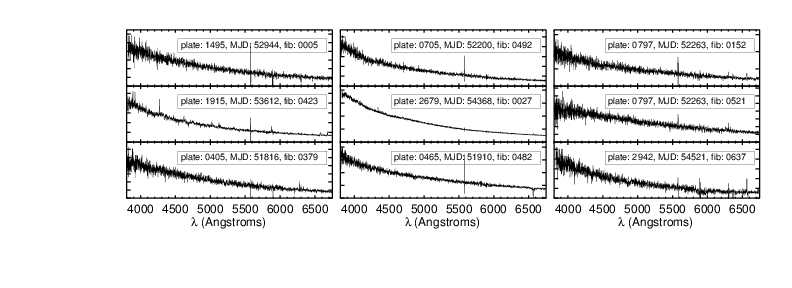}
\caption{Spectra of stars classified as ``Feat'' in Table 5. Figures 28.1 - 28.3, showing the entire set of spectra, are available in the online version of the Journal.}
\end{figure}

\newpage

\begin{deluxetable}{l}
\tablecaption{Query}
\startdata
\hline
\hline
SELECT\\
sp.ra, sp.dec, st.l, st.b, st.objId, pl.plate, pl.mjd, sp.fiberID, sp.specobjid,\\
sp.elodieLogG, sp.elodieFeH, sp.elodieSpType, sp.snMedian,\\
pl.plateQuality, sp.class, sp.subClass, sp.zWarning, st.psfmag\_g-st.extinction\_g
g0,\\
st.psfmag\_u-st.psfmag\_g-st.extinction\_u+st.extinction\_g umg0,\\
st.psfmag\_g-st.psfmag\_r-st.extinction\_g+st.extinction\_r gmr0\\
FROM\\
star st, specobj sp, plateX pl\\
WHERE\\
st.objID=sp.bestObjID\\
AND sp.plateID=pl.plateID\\
AND st.psfmag\_g-st.psfmag\_r -st.extinction\_g+st.extinction\_r $<$ -0.25\\
AND st.psfmag\_g $>$-9999\\
AND st.err\_g $>$ -1000\\
AND st.psfmag\_g-st.extinction\_g $<$22\\
AND st.psfmag\_r-st.extinction\_r $<$22\\
AND st.psfmag\_u-st.extinction\_u $<$22\\ 
AND st.extinction\_g $<$2\\
AND st.extinction\_u $<$2\\
AND st.extinction\_r $<$2\\
AND st.clean=1\\
\hline
\enddata

\end{deluxetable}

\newpage

\begin{deluxetable}{rrlll}
\tabletypesize{\scriptsize}
\tablewidth{0pt}
\tablecaption{Flags for Non-blue Stars}
\tablehead{
\colhead{Plate}&\colhead{MJD}&
\colhead{FiberID}&\colhead{Subclass}&\colhead{Flags}}
\startdata
304&51609&43&F5&DEBLENDED  AT  EDGE STATIONARY\\
&&&&BINNED1 INTERP COSMIC  RAY CHILD\\
2713&54400&142&F9&DEBLENDED  AT  EDGE MOVED BINNED1\\
&&&& DEBLENDED  AS  PSF INTERP CHILD\\
2713&54400&159&F5&DEBLENDED  AT  EDGE MOVED BINNED1\\
&&&& CHILD\\
606&52365&181&M5&STATIONARY BINNED2 BINNED1\\
&&&&DEBLEND  PRUNED MANYPETRO NODEBLEND BLENDED\\
965&52438&229&M4&DEBLENDED  AT  EDGE STATIONARY\\
&&&&MOVED BINNED1 INTERP CHILD\\
1869&53327&314&M6&BINNED1 MANYPETRO\\
2608&54474&355&F5&STATIONARY BINNED1 MANYPETRO\\
2509&54180&416&F5&DEBLENDED  AT  EDGE MOVED BINNED1\\
&&&&INTERP CHILD\\
1972&53466&433&M1&PSF  FLUX  INTERP INTERP  CENTER STATIONARY\\
&&&& BINNED1 INTERP MANYPETRO\\
2075&53730&470&F9&STATIONARY BINNED1\\
2077&53846&581&F5&STATIONARY BINNED1 INTERP\\
2125&53795&583&F5&STATIONARY BINNED1 MANYPETRO\\
2300&53682&584&F5&DEBLENDED  AT  EDGE STATIONARY MOVED\\
&&&& BINNED1 INTERP CHILD\\
2071&53741&614&G2&DEBLEND  NOPEAK DEBLENDED  AT  EDGE STATIONARY\\
&&&& BINNED1 INTERP NOPETRO CHILD\\
\hline
\enddata
\end{deluxetable}

\newpage

\begin{deluxetable}{llllllllllllll}
\tablecolumns{14}
\tabletypesize{\scriptsize}
\tablecaption{Classifications for Blue Stars}
\tablehead{
\colhead{Orig Class}&\colhead{Total}&\colhead{LowSN}&\colhead{Mis}&
\colhead{DA}&\colhead{DB}&\colhead{DO}&\colhead{DZ}&\colhead{UnuWD}&
\colhead{Bin}&\colhead{CV}&\colhead{B}&\colhead{Hot}&\colhead{Feat}}
\startdata
O&286&0&66&19&4&25&0&0&0&0&0&18&0\\
OB&1067&0&219&0&199&2&3&3&2&1&0&9&0\\
B&740&0&166&59&0&0&0&0&0&0&62&45&0\\
A&1718&0&13&1&0&0&0&0&0&0&11&1&0\\
F&5&3&2&0&0&0&0&0&0&2&0&0&0\\
G&1&0&1&0&0&0&0&0&0&1&0&0&0\\
K&2&0&2&0&0&0&0&1&1&0&0&0&0\\
M&72&0&72&0&0&0&0&0&70&0&0&0&2\\
L&65&22&43&15&4&2&0&9&0&0&0&1&12\\
T&52&15&37&13&6&0&0&5&0&0&0&0&13\\
CV&26&4&15&2&4&1&1&6&0&0&0&0&01\\
Carbon&1&0&1&1&0&0&0&0&0&0&0&0&0\\
Carbon WD&10&0&1&0&0&1&0&0&0&0&0&0&0\\
WD&7316&0&70&8&0&0&0&3&55&1&1&0&2\\
WD Magnetic&58&0&0&0&0&0&0&0&0&0&0&0&0\\
QSO&522&81&262&68&70&4&3&74&0&0&0&0&43\\
Galaxy&119&45&64&26&10&1&0&6&1&1&0&0&19\\
\hline
Total&12060&170&1034&212&297&36&7&107&129&6&74&74&92\\
\hline

\enddata

\end{deluxetable}

\newpage

\newpage

\begin{landscape}
\begin{deluxetable}{rrrrrlrlrrrrrllll}
\tabletypesize{\tiny}
\tablecolumns{16}
\tablewidth{0pt}
\tablecaption{Blue Stars that Match Templates}
\tablehead{
\colhead{RA}&\colhead{DEC}&\colhead{Plate}&\colhead{MJD}&
\colhead{Fiber}&\colhead{Class}&\colhead{Subclass}&\colhead{S$/$N}&\colhead{$g_0$}&
\colhead{$(u-g)_0$}&\colhead{$(g-r)_0$}&\colhead{$\mu_b$}&\colhead{$\mu_l \cos{b}$}&
\colhead{Kleinman}&\colhead{Eisenstein}&\colhead{Szkody}\\

\colhead{(deg.)} & \colhead{(deg.)} & & & & & & & & & & \colhead{(mas~yr$^{-1}$)} & \colhead{(mas~yr$^{-1}$)} & & & &
}
\startdata
  159.1486 & -0.010115702 & 274 & 51913 & 303 & STAR & WD & 14.7 & 18.6 & 0.31 & -0.29 & 9.5 & -41.3 & DA & DA & \\
  147.61614 & 0.093248577 & 267 & 51608 & 271 & STAR & OB & 13.2 & 18.27 & -0.16 & -0.41 & -0.8 & -9.7 & DB & DB & \\
  147.67551 & 1.2573303 & 267 & 51608 & 417 & STAR & WD & 19.7 & 17.67 & -0.03 & -0.44 & -14 & -44.8 & DA & DA & \\
  216.22703 & 3.0548088 & 584 & 52049 & 91 & STAR & WD & 4.8 & 19.61 & 0.18 & -0.3 & 5.6 & 13.5 & DA & DA & \\
  216.25333 & 3.6172128 & 584 & 52049 & 106 & STAR & WD & 3.2 & 20.17 & 0.18 & -0.32 & 0 & 0 & DA & DA & \\
  214.77188 & 4.4523433 & 584 & 52049 & 392 & STAR & WD & 25.1 & 17.22 & 0.16 & -0.34 & -24 & -9 & DA & DA & \\
  214.79398 & 4.525795 & 584 & 52049 & 395 & STAR & WD & 17.4 & 17.9 & 0.29 & -0.29 & -43 & 32.1 & DA & DA & \\
  215.98258 & 4.7655131 & 584 & 52049 & 515 & STAR & A0p & 5.7 & 19.36 & 1.05 & -0.28 & 5.9 & -0.7 &  & SD\_auto & \\
  165.22315 & 3.772999 & 580 & 52368 & 100 & STAR & O & 29.9 & 16.79 & -0.48 & -0.47 & -6.7 & -14.3 &  & SDO & \\
  165.68438 & 3.7030043 & 580 & 52368 & 47 & STAR & WD & 14.1 & 18.17 & 0.18 & -0.31 & 3.4 & -1.7 & DA & DA\_auto &

\enddata

\tablecomments{Table 4 is published in its entirety in the electronic edition of the Journal. A portion is shown here for guidance regarding its form and content.}
\end{deluxetable}
\end{landscape}

\newpage


\begin{landscape}
\begin{deluxetable}{rrrrrlrlrrrrrllll}
\tabletypesize{\tiny}
\tablecolumns{17}
\tablewidth{0pt}
\tablecaption{Reclassified Blue Stars}
\tablehead{
\colhead{RA}&\colhead{DEC}&\colhead{Plate}&\colhead{MJD}&
\colhead{Fiber}&\colhead{Template}&\colhead{S$/$N}&\colhead{$g_0$}&
\colhead{$(u-g)_0$}&\colhead{$(g-r)_0$}&\colhead{$\mu_b$}&\colhead{$\mu_l \cos{b}$}&
\colhead{$\Delta r_{\rm max}$}&\colhead{NewCl}&\colhead{Kleinman}&\colhead{Eisenstein}&\colhead{Szkody}\\

\colhead{(deg.)} & \colhead{(deg.)} & & & & & & & & & \colhead{(mas~yr$^{-1}$)} & \colhead{(mas~yr$^{-1}$)} & \colhead{(mags)} & & & &
}
\startdata
  2.536042 & -0.311611 & 686 & 52519 & 102 & L9 & 2.3 & 20.3 & -0.16 & -0.29 & 14.0 & -11.0 & 0.169(16) & LowSN & DB &  & \\
  35.292211 & -1.11163 & 2635 & 54114 & 249 & Galaxy & 1.1 & 21.61 & 0.27 & -0.32 & 0.0 & 0.0 & 0.659(36) & LowSN &  &  & \\
  37.278031 & -0.506203 & 705 & 52200 & 236 & QSO & 2.7 & 20.29 & -0.47 & -0.47 & 0.0 & 0.0 & 1.893(36) & LowSN &  & SDO: & \\
  37.570535 & -1.164097 & 705 & 52200 & 139 & L9 & 2.5 & 20.33 & -0.07 & -0.32 & 6.0 & 11.0 & 0.301(23) & LowSN & DB & DB & \\
  39.056296 & -0.40202 & 408 & 51821 & 227 & Galaxy & 2.9 & 20.09 & -0.49 & -0.45 & 2.0 & -1.0 & 0.184(25) & LowSN & DB:Z & DB: & \\
  47.606128 & 37.700939 & 2443 & 54082 & 39 & QSO & 4.5 & 19.82 & -0.36 & -0.41 & 11.0 & 0.0 & 0.000(2) & LowSN & DAH &  & \\
  48.136845 & 6.012435 & 2340 & 53733 & 467 & QSO & 3.4 & 20.09 & -0.56 & -0.56 & 0.0 & 0.0 & 0.000(1) & LowSN &  &  & \\
  57.01077 & -1.262932 & 1529 & 52930 & 262 & QSO & 2.3 & 19.88 & -0.58 & -0.52 & 0.0 & 0.0 & 0.910(30) & LowSN & DAH &  & \\
  58.92617 & 10.781151 & 3121 & 54749 & 94 & Galaxy & 5.0 & 20.05 & 1.06 & -0.44 & 0.0 & 0.0 & 0.000(1) & LowSN &  &  & \\
  70.722854 & 10.864697 & 2673 & 54096 & 197 & QSO & 1.2 & 19.93 & -0.03 & -0.34 & 0.0 & 0.0 & 0.000(1) & LowSN & DA &  &

\enddata

\tablecomments{Table 5 is published in its entirety in the electronic edition of the Journal. A portion is shown here for guidance regarding its form and content.}
\end{deluxetable}
\end{landscape}

%


\newpage

\begin{thebibliography}{}
\bibitem[Aihara et al.(2011)]{2011ApJS..193...29A} Aihara, H., Allende 
Prieto, C., An, D., et al.\ 2011, \apjs, 193, 29 
\bibitem[Caballero-Nieves et al.(2007)]{2007AJ....134.1072C} Caballero-Nieves, S.~M., Sowell, J.~R., \& Houk, N.\ 2007, \aj, 134, 1072
\bibitem[Dufour et al.(2007)]{2007ApJ...663.1291D} Dufour, P., Bergeron, 
P., Liebert, J., et al.\ 2007, \apj, 663, 1291 
\bibitem[Eisenstein et al.(2006)]{2006ApJS..167...40E} Eisenstein, D.~J., Liebert, J., Harris, H.~C., et al.\ 2006, \apjs, 167, 40 
\bibitem[Girven et al.(2011)]{2011MNRAS.417.1210G} Girven, J., 
G{\"a}nsicke, B.~T., Steeghs, D., \& Koester, D.\ 2011, \mnras, 417, 1210 
\bibitem[Gray \& Corbally(2009)]{graycorbally} Gray, R. O., \& Corbally, C. J. 2009, Stellar Spectral Classification, Princeton University Press, ISBN: 978-0-691-12511-4
\bibitem[Harris et al.(2003)]{2003AJ....126.1023H} Harris, H.~C., Liebert, J., Kleinman, S.~J., et al.\ 2003, \aj, 126, 1023 
\bibitem[Heber(2009)]{2009ARA&A..47..211H} Heber, U.\ 2009, \araa, 47, 211 
\bibitem[Heller et al.(2009)]{2009A&A...496..191H} Heller, R., Homeier, D., Dreizler, S., \& {\O}stensen, R.\ 2009, \aap, 496, 191 
\bibitem[Howell et al.(2013)]{2013AJ....145..109H} Howell, S.~B., Everett, M.~E., Seebode, S.~A., et al.\ 2013, \aj, 145, 109 
\bibitem[Ivezi{\'c} et al.(2002)]{ivezicetal02} Ivezi{\'c}, {\v Z}., Lupton, R.~H., Juri{\'c}, M., et al.\ 2002, \aj, 124, 2943 
\bibitem[Kleinman et al.(2013)]{2013yCat..22040005K} Kleinman, S.~J., 
Kepler, S.~O., Koester, D., et al.\ 2013, VizieR Online Data Catalog, 220, 40005 
\bibitem[Kramida et al.(2013)]{NIST} Kramida, A., Ralchenko, Yu., Reader, J. and NIST ASD Team (2013). NIST Atomic Spectra Database (version 5.1), [Online]. Available: http://physics.nist.gov/asd [Friday, 13-Jun-2014 23:24:54 EDT]. National Institute of Standards and Technology, Gaithersburg, MD.
\bibitem[Krzesi{\'n}ski et al.(2004)]{2004A&A...417.1093K} Krzesi{\'n}ski, J., Nitta, A., Kleinman, S.~J., et al.\ 2004, \aap, 417, 1093 
\bibitem[Lee et al.(2008)]{SSPP} Lee, Y.~S., Beers, T.~C., Sivarani, T., et al.\ 2008, \aj, 136, 2050 
\bibitem[Liebert et al.(2004)]{2004ApJ...606L.147L} Liebert, J., Bergeron, P., Eisenstein, D., et al.\ 2004, \apjl, 606, L147 
\bibitem[McCook \& Sion(2003)]{2003yCat.3235....0M} McCook, G.~P., \& Sion, E.~M.\ 2003, VizieR Online Data Catalog, 3235, 0 
\bibitem[McCook 
\& Sion(1999)]{1999ApJS..121....1M} McCook, G.~P., \& Sion, E.~M.\ 1999, \apjs, 121, 1 
\bibitem[Morgan \& Keenan(1973)]{1973ARA&A..11...29M} Morgan, W.~W., \& Keenan, P.~C.\ 1973, \araa, 11, 29 
\bibitem[Newberg et al.(2002)]{nyetal02} Newberg, H.~J., Yanny, B., Rockosi, C., et al.\ 2002, \apj, 569, 245 
\bibitem[Schlegel et al.(1998)]{sfd} Schlegel, D.~J., Finkbeiner, D.~P., \& Davis, M.\ 1998, \apj, 500, 525 
\bibitem[Shafter et al.(2011)]{2011ApJ...734...12S} Shafter, A.~W., Darnley, M.~J., Hornoch, K., et al.\ 2011, \apj, 734, 12 
\bibitem[Silvestri et al.(2006)]{2006AJ....131.1674S} Silvestri, N.~M., Hawley, S.~L., West, A.~A., et al.\ 2006, \aj, 131, 1674 
\bibitem[Sion et al.(1990)]{1990ApJS...72..707S} Sion, E.~M., Kenyon, 
S.~J., \& Aannestad, P.~A.\ 1990, \apjs, 72, 707 
\bibitem[Smolinski et al.(2011)]{2011AJ....141...89S} Smolinski, J.~P., 
Lee, Y.~S., Beers, T.~C., et al.\ 2011, \aj, 141, 89 
\bibitem[Solheim(2010)]{2010PASP..122.1133S} Solheim, J.-E.\ 2010, \pasp, 122, 1133 
\bibitem[Strauss et al.(1999)]{1999ApJ...522L..61S} Strauss, M.~A., Fan, X., Gunn, J.~E., et al.\ 1999, \apjl, 522, L61 
\bibitem[Szkody et al.(2011)]{2011AJ....142..181S} Szkody, P., Anderson, S.~F., Brooks, K., et al.\ 2011, \aj, 142, 181 
\bibitem[Valenti \& Piskunov(1996)]{1996A&AS..118..595V} Valenti, J.~A., \& Piskunov, N.\ 1996, \aaps, 118, 595 
\bibitem[Yanny et al.(2009)]{2009AJ....137.4377Y} Yanny, B., Rockosi, C., Newberg, H.~J., et al.\ 2009, \aj, 137, 4377 
\bibitem[Yanny et al.(2000)]{2000ApJ...540..825Y} Yanny, B., Newberg, 
H.~J., Kent, S., et al.\ 2000, \apj, 540, 825 
\bibitem[York et al.(2000)]{2000AJ....120.1579Y} York, D.~G., Adelman, J., Anderson, J.~E., Jr., et al.\ 2000, \aj, 120, 1579 

\end{thebibliography}
\end{document}